\documentclass[11pt]{article}
\usepackage{verbatim,amsmath,amssymb}
\usepackage{epsfig,float,color}
\usepackage{geometry}
\usepackage{setspace}
\usepackage{natbib}
\newcommand {\eqb}[1]{\begin{equation}\begin{array}{#1}}
\newcommand {\eqe}{\end{array}\end{equation}}

\newcommand {\esb}[1]{\begin{equation*}\begin{array}{#1}}
\newcommand {\ese}{\end{array}\end{equation*}}
\newcommand {\ds}{\displaystyle}

\newcommand {\pa}[2]{\frac{\partial{#1}}{\partial{#2}}}

\newcommand {\back}{\! \! \!}
\newcommand {\is}{\back &=& \back}
\newcommand {\dis}{\back &:=& \back}

\newcommand {\ais}{\back &\approx& \back}

\newcommand {\mi}{\back &-& \back}

\newcommand {\norm}[1]{\|#1\|}

\newcommand {\dif}{\mathrm{d}}

\newcommand {\II}{{I\kern-.3em I}}
\newcommand {\III}{{I\kern-.3em I\kern-.3em I}}

\newcommand {\mra}{\mathrm{a}}

\newcommand {\mrc}{\mathrm{c}}

\newcommand {\mrh}{\mathrm{h}}

\newcommand {\mrn}{\mathrm{n}}

\newcommand {\mrp}{\mathrm{p}}

\newcommand {\mrv}{\mathrm{v}}

\newcommand {\mf}{\mathbf{f}}

\newcommand {\mh}{\mathbf{h}}

\newcommand {\mk}{\mathbf{k}}
\newcommand {\ml}{\mathbf{l}}

\newcommand {\mw}{\mathbf{w}}
\newcommand {\mx}{\mathbf{x}}

\newcommand {\ba}{\boldsymbol{a}}
\newcommand {\bb}{\boldsymbol{b}}

\newcommand {\bff}{\boldsymbol{f}}
\newcommand {\bg}{\boldsymbol{g}}

\newcommand {\bm}{\boldsymbol{m}}
\newcommand {\bn}{\boldsymbol{n}}

\newcommand {\bt}{\boldsymbol{t}}
\newcommand {\bu}{\boldsymbol{u}}
\newcommand {\bv}{\boldsymbol{v}}
\newcommand {\bw}{\boldsymbol{w}}
\newcommand {\bx}{\boldsymbol{x}}
\newcommand {\by}{\boldsymbol{y}}

\newcommand {\mN}{\mathbf{N}}

\newcommand {\mX}{\mathbf{X}}

\newcommand {\bA}{\boldsymbol{A}}
\newcommand {\bB}{\boldsymbol{B}}
\newcommand {\bC}{\boldsymbol{C}}

\newcommand {\bF}{\boldsymbol{F}}

\newcommand {\bN}{\boldsymbol{N}}

\newcommand {\bX}{\boldsymbol{X}}

\newcommand {\sig}{\sigma}

\newcommand {\bsig}{\mbox{\boldmath$\sigma$}}

\newcommand {\btau}{\mbox{\boldmath$\tau$}}

\newcommand {\bone}{\mathbf{1}}

\newcommand {\bbR}{\mathbb{R}}

\newcommand {\IR}{{\rm\kern.24em
   \vrule width.02em height1.53ex depth-.05ex
   \kern-.3em R}}
\newcommand {\ic}{{\rm\kern.20em
   \vrule width.02em height1.0ex depth-.05ex
   \kern-.22em c}}
\newcommand {\ia}{{\rm\kern.20em
   \vrule width.02em height1.05ex depth-.0ex
   \kern-.25em a}}
\newcommand {\IC}{{\rm\kern.24em
   \vrule width.02em height1.4ex depth-.05ex
   \kern-.26em C}}
\newcommand {\ID}{{\rm\kern.34em
   \vrule width.02em height1.5ex depth-.05ex
   \kern-.36em D}}
\newcommand {\IS}{{\rm\kern.24em
   \vrule width.02em height1.6ex depth.05ex
   \kern-.26em S}}
\newcommand {\IT}{{\rm\kern.50em
   \vrule width.02em height1.55ex depth-.05ex
   \kern-.52em T}}

\newcommand {\IE}{{\rm\kern.24em
   \vrule width.02em height1.55ex depth-.05ex
   \kern-.33em E}}
\newcommand {\IEa}{{\rm\kern.24em
   \vrule width.02em height1.55ex depth-.05ex
   \kern-.33em E}^{1}_{ijkl}}
\newcommand {\IEb}{{\rm\kern.24em
   \vrule width.02em height1.55ex depth-.05ex
   \kern-.33em E}^{2}_{ijkl}}

\newcommand {\sD}{\mathcal{D}}

\newcommand {\sP}{\mathcal{P}}

\newcommand {\sS}{\mathcal{S}}

\newcommand {\sW}{\mathcal{W}}

\newcommand {\Ass}[2]{\kern 0.9ex \vrule width0.45em height0.2ex depth0ex \kern -2.1ex \bigwedge_{#1}^{#2}}
\newcommand {\ASS}[2]{\kern 1.45ex \vrule width0.5em height0.2ex depth0ex \kern -2.65ex \bigwedge_{#1}^{#2}}

\begin{document}

\begin{center}
\Large{\bf{A computational formulation for constrained solid and liquid membranes 
considering isogeometric finite elements}}\\

\end{center}

\begin{center}
\large{Roger A. Sauer
\footnote{corresponding author, email: sauer@aices.rwth-aachen.de}, Thang X. Duong, Callum J. Corbett}\\
\vspace{4mm}

\small{\textit{Aachen Institute for Advanced Study in Computational Engineering Science (AICES), RWTH Aachen
University, Templergaben 55, 52056 Aachen, Germany}}

\vspace{4mm}

Submitted on October 17$^\mathrm{th}$ 2012 for publication in \\
\textit{Computer Methods in Applied Mechanics and Engineering}

\end{center}

\vspace{3mm}


\rule{\linewidth}{.15mm}
{\bf Abstract}

A geometrically exact membrane formulation is presented that is based on curvilinear coordinates and isogeometric finite elements, and is suitable for both solid and liquid membranes. The curvilinear coordinate system is used to describe both the theory and the finite element equations of the membrane. In the latter case this avoids the use of local cartesian coordinates at the element level. Consequently, no transformation of derivatives is required. The formulation considers a split of the in-plane and out-of-plane membrane contributions, which allows the construction of a stable formulation for liquid membranes with constant surface tension.
The proposed membrane formulation is general, and accounts for dead and live loading, as well as enclosed volume, area, and contact constraints. 
The new formulation is illustrated by several challenging examples, considering linear and quadratic Lagrange elements, as well as isogeometric elements based on quadratic NURBS and cubic T-splines. It is seen that the isogeometric elements are much more accurate than standard Lagrange elements. The gain is especially large for the liquid membrane formulation since it depends explicitly on the surface curvature.


\vspace{-4mm}
\rule{\linewidth}{.15mm}


\section{Introduction}\label{s:intro}

Membranes are computationally challenging structures. Their geometry can be complex, they may undergo large deformations, large rotations and large strains - and thereby behave highly nonlinear - and they are characterized by several physical instabilities: They are unstable in compression, unstable for out-of-plane loading (in the case of zero in-plane tension), 
unstable for pressure loading (in the case of rubber membranes) and unstable w.r.t. in-plane loading (in the case of liquid membranes).
The aim of this paper is to formulate a general, 3D, geometrically exact and fully nonlinear membrane model 
that accounts for pressure loading as well as volume, area, and contact constraints
and is suitable for both solid and liquid membranes.
Our focus is on pure membranes, i.e. curved, surface structures that do not support in-plane compression, out-of-plane bending, and shear.\footnote{We note that in the literature, the term membrane is often also used for the special case of 2D plane-stress structures.}
Such a restricted focus is useful due to the large range of membrane applications:
they appear
as inflatable and pressurized structures, like balloons, tubes and airbags;
as fabrics, tents, canopies, parachutes, foils and sails;
as water-filled membrane structures, like inflatable dams;
as biological membranes, like blood vessels, cell, diaphragms, aneurysms and lung alveoli;
as liquid droplets, menisci, bubbles, foams and sprays; 
as thin sheets and films - both liquid and solid -
as atomistic membranes, like graphene sheets;
as interacting membranes, e.g. adhering cells; and
in the topic of form-finding and minimal surfaces.

Computational formulations for 3D, nonlinear membrane go back to the seminal work of Oden (\citet{oden67}, see also \citet{oden}). Since then, the field has been continuously advanced, among others by 
\citet{fried82,tang82,roddeman87b,contri88,wriggers90,ibrahim93,haseganu94,gosling96a,muttin96,wu96,bonet00,rumpel03,stanuszek03,weinberg08}.
Many of these works are concerned with the topic of wrinkling due to in-plane compression.
More recently, computational formulations based on curvilinear coordinates have been considered rigorously, both for
membranes \citep{ambroziak06} 
and shells \citep{arciniega07}.
Another recent development are
rotation-free shell formulations, 
as they have been considered by
\citet{flores07,linhard07,dung08} and recently \citet{benson11,thanh11} for isogeometric analysis.
%
Isogeometric formulations allow the formulation of $C^1$-continuous surface formulations that are advantageous for flow simulations \citep{kiendl10} and sliding contact \citep{lorenzis11,temizer12}, see also \citet{sauer10b,sauer-ece2} for Hermite-based, $C^1$-continuous contact surfaces.
Relevant to membranes is also the topic of live pressure loading \citep{bufler84,schweizerhof84}. 
Membranes are also an interesting subject in shape optimization \citep{bletzinger05,manh11}.

The presented formulation contains several merits and novelties:
It allows a split between in-plane and out-of-plane contributions, which is used to construct a new formulation for liquid membranes.
It admits arbitrary elastic material models for solid and liquid membranes.
It is based purely on displacement-based finite elements and can be used with any kind of such elements. 
It includes, in particular, isogeometric NURBS elements to capture the deforming surface geometry to high-accuracy, even for comparably coarse discretizations.
It is straight forward to implement in an existing FE framework.
It avoids the need to transform derivatives between configurations and avoids the use of local cartesian coordinate systems.
Shells models are often formulated using local cartesian coordinate systems, as this allows using classical constitutive relations formulated in this manner \citep{wriggers-fee}.
To our mind, there is no need for such a detour: The balance laws, kinematics, constitutive relations as well as the FE weak forms and corresponding FE arrays can all be formulated efficiently in the curvilinear coordinate system.
The capabilities of the presented formulation are demonstrated by several challenging computational examples, considering pressure loading, inflation and contact.


The following section presents the theory of nonlinear membranes in the framework of curvilinear coordinates, considering pressure loading, volume, area and contact constraints. Sec.~\ref{s:fed} proposes a straight-forward finite element implementation of the theory, and Sec.~\ref{s:ne} presents several examples of solid and liquid membranes to illustrate the capabilities of the present formulation.

\section{Nonlinear membranes}\label{s:membrane}

In this section, we summarize the theory of nonlinear membranes in the framework of curvilinear coordinates. The membrane kinematics, constitution and balance laws in strong and weak form are discussed, and various kinds of constraints are considered.

\subsection{Surface description in curvilinear coordinates}\label{s:surface}

The membrane surface, denoted $\sS$, is fully characterized by the parametric description
\eqb{l}
\bx = \bx\big(\xi^1,\xi^2\big)~.
\eqe
%
This corresponds to a mapping of the point $(\xi^1,\xi^2)$ in the parameter domain $\sP$ to the material point $\bx\in\sS$. 
In the following, Greek letters are used to denote the two indices 1 and 2. Summation is then implied on repeated indices. 
The tangent vectors to coordinate $\xi^\alpha$ at 
point $\bx\in\sS$ are given by
\eqb{l}
\ba_{\alpha} = \ds\pa{\bx}{\xi^{\alpha}}~,\quad \alpha=1,2
\label{e:a_a}\eqe
The two vectors form a basis for the tangent plane of $\sS$. In general, they are not orthonormal. This apparent drawback of the description is actually an advantage when it comes to the kinematical description. This turns out to be very straightforward, e.g. see Eq.~(\ref{e:F}). The basis at $\bx$ is characterized by the metric tensor, that has the
co-variant components 
\eqb{l}
a_{\alpha\beta} := \ba_\alpha\cdot\ba_\beta~.
\label{e:aab}\eqe
From the inversion
\eqb{l}
[a^{\alpha\beta}] := [a_{\alpha\beta}]^{-1}
\eqe
we obtain the contra-variant components of the metric tensor.
Explicitly, we have $a^{11}=a_{11}/\det a_{\alpha\beta}$, $a^{12}=-a_{12}/\det a_{\alpha\beta}$ and $a^{22}=a_{22}/\det a_{\alpha\beta}$.
With these a dual basis can be constructed: From the co-variant base vectors $\ba_\alpha$ the contra-variant counterparts
\eqb{l}
\ba^\alpha := a^{\alpha\beta}\ba_\beta
\label{e:aa}\eqe
can be determined. We note that summation is implied on repeated indices. Note that $\ba^\alpha\cdot\ba^\beta=a^{\alpha\beta}$ and $\ba_\alpha\cdot\ba^\beta=\delta_\alpha^\beta$, where $\delta_\alpha^\beta$ is the Kronecker symbol.
The unit normal of $\sS$ at $\bx$ is given by
\eqb{l}
\bn = \ds\frac{\ba_1\times\ba_2}{\norm{\ba_1\times\ba_2}}~.
\label{e:n}\eqe
It can be shown that 
\eqb{l}
\norm{\ba_1\times\ba_2}=\sqrt{\det a_{\alpha\beta}}~. 
\label{e:naa}\eqe
The bases $\{\ba_1,\ba_2,\bn\}$ and $\{\ba^1,\ba^2,\bn\}$ can then be used to decompose any vector $\bv$ on $\sS$, i.e.
\eqb{l}
\bv=v^\alpha\,\ba_\alpha + v_\mrn\,\bn =v_\alpha\,\ba^\alpha + v_\mrn\,\bn~,
\label{e:vector}\eqe
where $v_\alpha$ denote the co-variant, and $v^\alpha$ the contra-variant components of $\bv$.
The derivative of the tangent vectors is given by
\eqb{l}
\ba_{\alpha,\beta}=\ds\frac{\partial\ba_{\alpha}}{\partial\xi^{\beta}}~.
\eqe
Further, we require the so-called co-variant derivative of $\ba_\alpha$, which is defined by
\eqb{l}
\ba_{\alpha;\beta}:=\ba_{\alpha,\beta}-\Gamma_{\alpha\beta}^{\gamma}\,\ba_{\gamma}
\eqe
where $\Gamma_{\alpha\beta}^{\gamma}$ are the Christoffel symbols of the second kind given by $\Gamma_{\alpha\beta}^{\gamma}=\ba_{\alpha,\beta}\cdot\ba^\gamma$.
Introducing the identity tensor on $\sS$
\eqb{l}
\bone =  \ba_{\alpha}\otimes\ba^{\alpha} = \ba^{\alpha}\otimes\ba_{\alpha} = \tilde\bone-\bn\otimes\bn~,
\eqe
where $\tilde\bone$ is the usual identity tensor in $\bbR^3$, \footnote{A tilde is used here to indicate standard tensors in $\bbR^3$} we can write
\eqb{l}
\ba_{\alpha;\beta}=(\bn\otimes\bn)\,\ba_{\alpha,\beta}~.
\eqe
Contracting with $\bn$ then yields
\eqb{l}
\bn\cdot\ba_{\alpha;\beta}=\bn\cdot\ba_{\alpha,\beta} = b_{\alpha\beta} 
\label{e:b_ab}\eqe
which are the co-variant components of the curvature tensor $\bb=b_{\alpha\beta}\,\ba^\alpha\otimes\ba^\beta$. The eigenvalues of this tensor are the principal curvatures of surface $\sS$.

\subsection{Membrane kinematics}

\begin{figure}[h]
\begin{center} \unitlength1cm
\includegraphics[height=80mm]{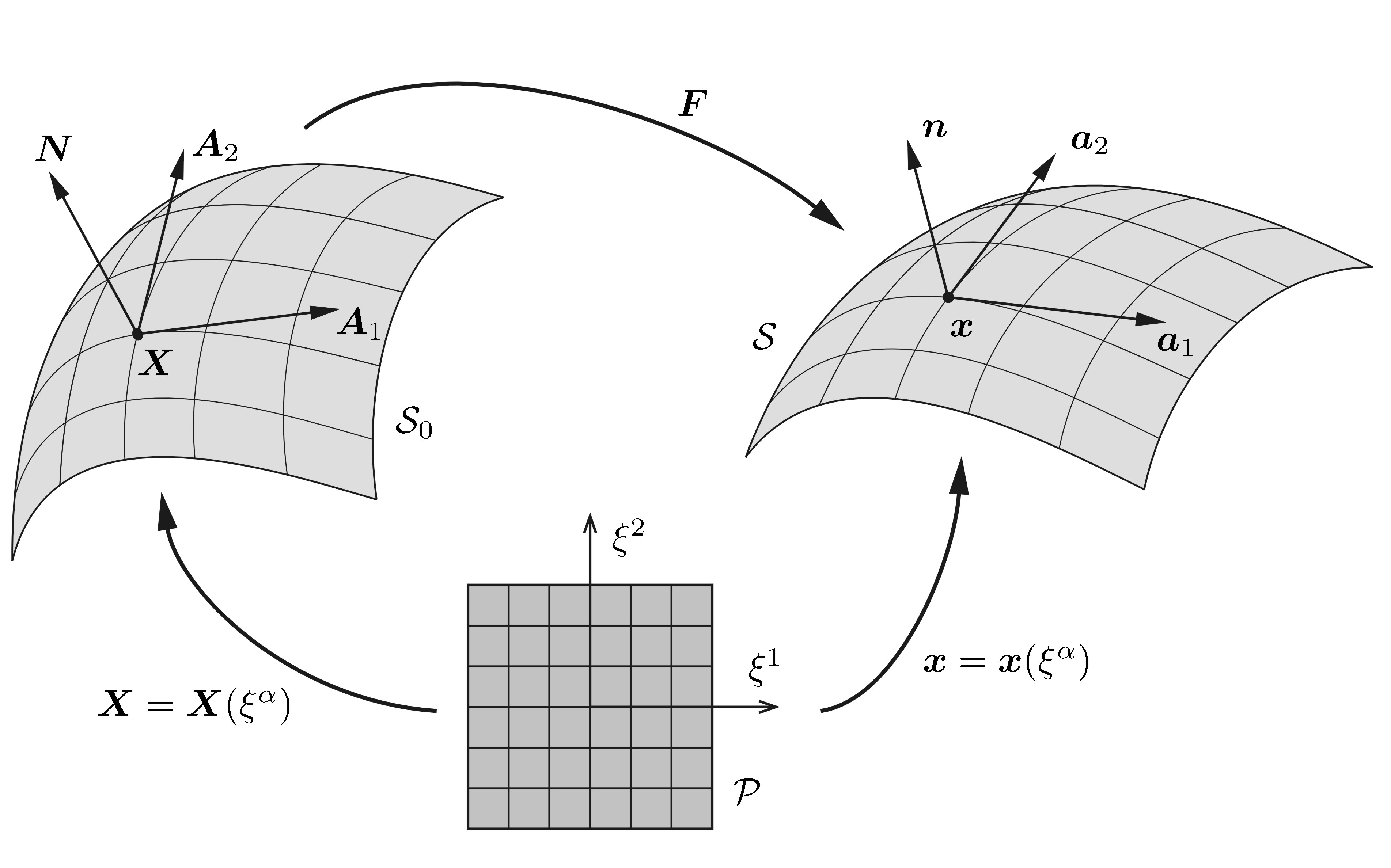}
\caption{Mapping between parameter domain $\sP$, reference surface $\sS_0$ and current surface $\sS$ }
\label{f:mapping}
\end{center}
\end{figure}

Next, we consider the deformation of the membrane surface. We therefore distinguish between the deformed, current configuration $\sS$ and the undeformed, initial configuration $\sS_0$, see Fig.~\ref{f:mapping}.
Both surfaces are described by the relations of Sec.~\ref{s:surface}. For surface $\sS$ we use the lower case symbols $\bx$, $\ba_\alpha$, $a_{\alpha\beta}$, $\ba^\alpha$, $\bn$ and $b_{\alpha\beta}$. For surface $\sS_0$ we use the corresponding upper case symbols $\bX$, $\bA_\alpha$, $A_{\alpha\beta}$, $\bA^\alpha$ and $\bN$.\footnote{Here, the curvature tensor $[b_{\alpha\beta}]$ is only needed on $\sS$.}
In order to characterize the deformation between surfaces $\sS_0$ and $\sS$ consider the line element
\eqb{l}
\dif\bx = \ds\pa{\bx}{\xi^\alpha}\dif\xi^\alpha = \ba_\alpha\,\dif\xi^\alpha
\eqe
and likewise $\dif\bX = \bA_\alpha\,\dif\xi^\alpha$. Contracting with $\bA^\beta$ yields
$\dif\xi^\alpha=\bA^\alpha\cdot\dif\bX$, so that
\eqb{l}
\dif\bx = (\ba_\alpha\otimes\bA^\alpha)\,\dif\bX~.
\eqe
Here the tensor
\eqb{l}
\bF=\ba_{\alpha}\otimes\bA^{\alpha}
\label{e:F}\eqe
is the surface deformation gradient of the mapping $\bX\rightarrow\bx$. Likewise we find $\bF^{-1}=\bA_{\alpha}\otimes\ba^{\alpha}$. Through $\bF$ we thus have the following transformations
\eqb{llllll}
\ba_{\alpha} \is \bF\bA_{\alpha}~,& \bA_{\alpha} \is \bF^{-1}\ba_{\alpha}~, \\[1mm]
\ba^{\alpha} \is \bF^{-T}\bA^{\alpha}~,~& \bA^{\alpha} \is \bF^{T}\ba^{\alpha}~.
\eqe
Given $\bF$, we can introduce the right and left Cauchy-Green surface tensors and their inverses, i.e.
\eqb{llllll}
\bC \is \bF^T\bF=a_{\alpha\beta}\,\bA^{\alpha}\otimes\bA^{\beta}~,~ & \bC^{-1} 
=a^{\alpha\beta}\,\bA_{\alpha}\otimes\bA_{\beta}~, \\[1mm]
\bB \is \bF\bF^T=A^{\alpha\beta}\,\ba_{\alpha}\otimes\ba_{\beta}~,~ & \bB^{-1} 
=A_{\alpha\beta}\,\ba^{\alpha}\otimes\ba^{\beta}~.
\eqe

Next we discuss the surface stretch between surfaces $\sS_0$ and $\sS$. The area element $\dif a\subset\sS$ is defined by
\eqb{l}
\dif a := \norm{(\ba_1\,\dif\xi^1)\times(\ba_2\,\dif\xi^2)} = \norm{\ba_1\times\ba_2}\,\dif\Box~,
\eqe
where $\dif\Box:=\dif\xi^1\,\dif\xi^2$. A corresponding statement follows for $\dif A\subset\sS_0$. In view of Eq.~(\ref{e:naa}) we thus have the relations
\eqb{lllrll}
\dif A \is J_A\,\dif\Box~,~& J_A \dis \sqrt{\det A_{\alpha\beta}}~, \\[1mm]
\dif a \is J_a\,\dif\Box~,~& J_a \dis \sqrt{\det a_{\alpha\beta}}~, \\[1mm]
\dif a \is J\,\dif A~,~& J \dis J_a/J_A~.
\label{e:da}\eqe

\subsection{Momentum balance for membranes}

From the balance of linear momentum the strong form equilibrium equation
\eqb{l}
\bt^{\alpha}_{;\alpha} + \bff = \boldsymbol{0}~,
\label{e:sf}\eqe
at $\bx\in\sS$ can be obtained \citep{steigmann99}. Here $\bff$ is a distributed surface force, that can be decomposed as        
\eqb{l}
\bff = f_\alpha\,\ba^\alpha + p\,\bn = f^\alpha\,\ba_\alpha + p\,\bn~
\eqe
where $f_\alpha$ and $f^\alpha$ are the co-variant and contra-variant in-plane components of $\bff$ and
$p$ is the out-of-plane pressure acting on $\sS$. Further, $\bt^\alpha$ denotes the internal traction acting on the internal surface $\perp$ $\ba^\alpha$. According to Cauchy's formula
\eqb{ll}
\bt^\alpha = \bsig\,\ba^\alpha~,
\eqe
where $\bsig$ denotes the Cauchy stress tensor at $\bx\in\sS$, which we consider to be symmetric. We emphasize that $\bt^\alpha$ is not a physical traction 
since $\ba^\alpha$ is usually not normalized.
In general, the stress tensor takes the form
\eqb{l}
\bsig = \sigma^{\alpha\beta}\,\ba_\alpha\otimes\ba_\beta
+ \sigma^{3\alpha}(\bn\otimes\ba_\alpha+\ba_\alpha\otimes\bn)
+ \sigma^{33}\,\bn\otimes\bn~.
\eqe
For membranes it is typically assumed that $\sigma^{3\alpha}=\sigma^{33}=0$, so that
\esb{l}
\bsig = \sigma^{\alpha\beta}\,\ba_\alpha\otimes\ba_\beta~.
\ese
In this case we find that $\bt^\alpha=\sigma^{\beta\alpha}\ba_\beta$ such that the co-variant derivative of $\bt^\alpha$ becomes
\eqb{l}
\bt^\alpha_{;\alpha} = \sigma^{\alpha\beta}_{~~;\alpha}\,\ba_\beta + \sigma^{\beta\alpha}\,b_{\alpha\beta}\,\bn
\eqe
according to Eq.~(\ref{e:b_ab}). Equilibrium equation (\ref{e:sf}) thus decomposes into
\eqb{llll}
\sig^{\alpha\beta}_{~~;\alpha} + f^\beta \is 0~,~ & $(in-plane equilibrium)$, \\[1mm]
\sig^{\alpha\beta}\,b_{\alpha\beta} + p \is 0~,~ & $(out-of-plane equilibrium)$.
\label{e:dsf}\eqe
To close the problem, the usual Dirichlet and Neumann boundary conditions
\eqb{llll}
\bu \is \bar\bu & $on$~\partial_u\sS \\[1mm]
\bt \is \bar\bt & $on$~\partial_t\sS
\eqe
are considered on the membrane boundary $\partial\sS=\partial_u\sS\,\cup\,\partial_t\sS$. 
Here, we suppose that the prescribed traction $\bar\bt=\bar t^\alpha\,\ba_\alpha$ is tangent to $\sS$, since out-of-plane boundary forces, as well as out-of-plane line and point loads within the surface, lead to singularities in the membrane deformation and are therefore not considered in the present formulation. The traction on boundary $\partial_t\sS$, according to Cauchy's formula, is given by
\eqb{l}
\bt=\bsig\bm~, 
\eqe
where $\bm=m_\alpha\ba^\alpha$ is the outward unit normal of $\partial_t\sS$. It follows that $\bt=m_\alpha\bt^\alpha$.

\subsection{Membrane constitution}\label{s:consti}

The known 3D constitutive models can be adapted to the membrane. We therefore suppose a general elastic material relation of the form $\tilde\bsig=\tilde\bsig(\tilde\bB)$. For membranes it is useful to consider the decomposition
$\tilde\bB = \bB + \lambda^2_3\,(\bn\otimes\bn)$, where $\lambda_3$ is the out-of-plane stretch,
and $\tilde\bsig = \bsig/t + \sig_{33}\,(\bn\otimes\bn)$, where $\bsig$ is defined as the in-plane stress tensor (with units force per length) and $t=\lambda_3 T$ denotes the current membrane thickness, for a given reference thickness $T$.
Out of these considerations, a relation between the membrane quantities $\bB$ and $\bsig$ can be obtained.
As an example we consider an incompressible Neo-Hooke material, given by
\eqb{l}
\tilde\bsig = \tilde\mu\tilde\bB + q\tilde\bone~,
\label{e:incompNH}\eqe
where $\tilde\mu$ is the shear modulus and $q$ denotes the Lagrange multiplier associated with the incompressibility constraint. For membranes, the model decomposes into
\eqb{lll}
\bsig \is ( \tilde\mu\bB + q\bone)\,T/\lambda_3~,\\[1mm] 
\sig_{33} \is \tilde\mu\lambda_3^2 + q~.
\eqe
For incompressibility $\det\tilde\bB=(J\lambda_3)^2=1$. Under the plane stress assumption $\sig_{33}=0$, we then find $q=-\tilde\mu/J^2$ and consequently 
\eqb{l}
\bsig = \mu/J\big(\bB -\bone/J^2\big)~,
\label{e:incompMNH}\eqe
with $\mu:=\tilde\mu T$. Componentwise, in the $\ba_\alpha$ basis, this becomes
\eqb{l}
\sig^{\alpha\beta} = \mu/J\big(A^{\alpha\beta} -a^{\alpha\beta}/J^2\big)~.
\label{e:incompNH_ab}\eqe
Contracting with $a_{\beta\gamma}$, the components $\sig^\alpha_\beta$ and $\sig_{\alpha\beta}$ can be obtained.\footnote{Due to the symmetry of $\sig^{\alpha\beta}$ the ordering of indices does not matter in $\sig^{\alpha}_\beta$, i.e. $\sig^{\alpha}_{~\beta}=\sig^{~\alpha}_{\beta}$.}

Another example are liquid, e.g.~water, membranes governed by constant isotropic surface tension $\gamma$. In that case a constant stress tensor of the form 
\eqb{l}
\sig^\alpha_\beta=\gamma\,\delta^\alpha_\beta
\label{e:water}\eqe
is obtained. It can be seen that the in-plane equilibrium equation~(\ref{e:dsf}.1) is only satisfied for $f^\alpha=0$. This implies that static water membranes cannot equilibrate in-plane loads, and are therefore unstable in-plane; a property that needs to be addressed in a computational formulation (see Sec.~\ref{s:fed}). The out-of-plane equation~(\ref{e:dsf}.2) now yields
\eqb{l}
2H\gamma + p = 0~,\quad 2H:=b^\alpha_\alpha~,
\label{e:YL}\eqe
which is the well known Young-Laplace equation. A prominent feature of liquid membranes is that they form distinct contact angles. This property is not addressed here.

\subsection{Membrane weak form}\label{s:wf}

Next, we derive the weak form corresponding to equilibrium equation~(\ref{e:sf}).
Consider a kinematically admissible variation of $\sS$, denoted $\bw\in\sW$, where $\sW$ denotes a suitable space for $\bw$.
Contracting Eq.~(\ref{e:sf}) with $\bw$ and integrating over $\sS$ yields
\eqb{l}
\ds\int_{\sS}\bw\cdot\big(\bt^{\alpha}_{;\alpha}+\bff\big)\, \dif a =0\quad\forall\bw\in\sW~.
\label{e:wf_1}\eqe
Considering $\bw=w_\alpha\,\ba^\alpha+w\,\bn$, 
this expands into
\eqb{l}
\ds\int_{\sS}w_\alpha\big(\sig^{\alpha\beta}_{~~;\beta}+f^\alpha\big)\, \dif a 
+ \int_\sS w\,\big(\sig^{\alpha\beta}\,b_{\alpha\beta}+p\big)\,\dif a =0\quad\forall\bw\in\sW~,
\label{e:wf_2}\eqe
i.e. it splits into the in-plane and out-of-plane parts identified in Eq.~(\ref{e:dsf}). 
Such a split is useful if different approximation techniques are chosen for the in-plane and out-of-plane response.
Using the divergence theorem for curved surfaces \citep{gurtin75}, the first in-plane term is rewritten into
\eqb{lllll}
\ds\int_{\sS}w_\alpha\,\sig^{\alpha\beta}_{~~;\beta}\, \dif a
\is \ds\int_\sS\big(w_\alpha\,\sig^{\alpha\beta}\big)_{;\beta}\, \dif a \mi \ds\int_{\sS}w_{\alpha;\beta}\,\sig^{\alpha\beta}\, \dif a \\[4mm]
\is \ds\int_{\partial\sS}w_\alpha\,\sig^{\alpha\beta}\,m_\beta\, \dif s \mi \ds\int_{\sS}w_{\alpha;\beta}\,\sig^{\alpha\beta}\, \dif a~,
\eqe
where $m_\alpha=\bm\cdot\ba_\alpha$ are the co-variant components of the unit normal $\bm$ on the line $\partial\sS$.
Since $w^\alpha=0$ on $\partial_u\sS$ and since 
$w_\alpha\sig^{\alpha\beta}m_\beta 
= w_\alpha\,\bar t^\alpha$
on $\partial_t\sS$ expression (\ref{e:wf_2}) thus becomes
\eqb{l}
G_{\mathrm{int}} - G_{\mathrm{ext}}=0\quad\forall\bw\in\sW~,
\label{e:wf_3}\eqe
where
\eqb{lll}
G_\mathrm{int} \dis \ds\int_{\sS}w_{\alpha;\beta}\,\sig^{\alpha\beta}\, \dif a
- \int_\sS w\,\sig^{\alpha\beta}\,b_{\alpha\beta}\,\dif a~, \\[4mm]
G_\mathrm{ext} \dis \ds\int_\sS w_\alpha\,f^\alpha\,\dif a
+ \int_{\partial_t\sS}w_\alpha\,\bar t^\alpha\,\dif s
+ \int_\sS w\,p\,\dif a~,
\label{e:Gintext}\eqe
are the internal and external virtual work contribution due to variation $\bw$. Considering $w_\alpha=0$ and $w=0$ subsequently, the weak form can be decomposed into the 
weak forms
\eqb{rlll}
\ds\int_{\sS}w_{\alpha;\beta}\,\sig^{\alpha\beta}\, \dif a - \ds\int_\sS w_\alpha\,f^\alpha\,\dif a
- \int_{\partial_t\sS}w_\alpha\,\bar t^\alpha\,\dif s \is 0\quad \forall\,w_\alpha\in\sW_\alpha ~& $(in-plane),$ \\[4mm]
\ds\int_\sS w\,\sig^{\alpha\beta}\,b_{\alpha\beta}\,\dif a + \int_\sS w\,p\,\dif a \is 0 \quad \forall\,w\in\sW_\mrn & $(out-of plane).$
\label{e:wfs}\eqe
Such a split is advantageous for the description of liquid membranes. Since liquid membranes are inherently unstable in-plane, they can be stabilized by providing additional stiffness via Eq.~(\ref{e:wfs}.1) without affecting the out-of-plane response. This is demonstrated in the examples of Sec.~\ref{s:droplet} and \ref{s:wcontact}. \\
Otherwise, considering
\eqb{lll}
w_{\alpha;\beta} \is (\bw\cdot\ba_\alpha)_{;\beta} 
	= \bw_{;\beta}\cdot\ba_\alpha + w\,b_{\alpha\beta}~,
\label{e:w_ab}\eqe
the two terms of $G_\mathrm{int}$ can be combined into
\eqb{l}
G_{\mathrm{int}}= \ds\int_{\sS}\bw_{;\alpha}\cdot\sig^{\alpha\beta}\,\ba_\beta\, \dif a~.
\label{e:Gint1}\eqe
It is noted that for this expression only single derivatives of variation $\bw$ and configuration $\bx$ are required, while in the decomposed formulation 
of Eq.~(\ref{e:wfs}) second derivatives of $\bx$ appear.
Introducing the surface Kirchhoff stress tensor $\btau=J\bsig$, which eliminates one $J$ from expression~(\ref{e:incompNH_ab}), the last equation can be rewritten into
\eqb{l}
G_{\mathrm{int}}= \ds\int_{\sS_0}\bw_{;\alpha}\cdot\tau^{\alpha\beta}\,\ba_\beta\, \dif A~.
\label{e:Gint2}\eqe
Within framework (\ref{e:wf_3}), both dead and live loading can be considered. This is discussed further in Sec.~\ref{s:dwf}. Beforehand, we discuss several useful membrane constraints.

\subsection{Volume constraints}

The volume of the domain $\sD$ enclosed by the membrane may be constrained. An example is a cell containing incompressible fluid. Formally, the volume constraint is written as
\eqb{l}
g_\mrv := V-V_0 = 0~,
\label{e:gv}\eqe
where $V_0$ and $V$ denote the initial and current volumes enclosed by the initial and current membrane configurations. Since $\dif v=\bx\cdot\bn\,\dif a/3$ and $\dif V=\bX\cdot\bN\,\dif A/3$, these can be computed by the surface integration
\eqb{l}
V = 
\ds\frac{1}{3}\int_{\sS}\bx\cdot\bn\, \dif a~,\quad
V_0 = 
\ds\frac{1}{3}\int_{\sS_0}\bX\cdot\bN\, \dif A
\label{e:V}\eqe
These expressions are valid for closed surfaces, and care has to be taken when modeling open membranes. In this case one must account for the volume contribution associated with the missing surface. Eq.~(\ref{e:gv}) can be 
included in the formulation by the Lagrange multiplier method. The Lagrange multiplier associated with the volume constraint is the internal membrane pressure $p$.

Remark: We note that the governing equations (\ref{e:wf_3}) and (\ref{e:gv}) can be derived from a variational principle for conservative systems. This is the case for the constitutive models discussed in Sec.~\ref{s:consti} and for pressure loading of closed membranes surfaces.

\subsection{Area constraints}

Another useful constraint is a constraint on the membrane surface area. For example, red blood cells are known to conserve the surface area during deformation \citep{kloeppel11}. Formally this is expressed as 
\eqb{l}
g_\mra := A - A_0 = 0~,
\eqe
with
\eqb{l}
A = \ds\int_{\sS}\dif a~,\quad
A_0 = \ds\int_{\sS_0} \dif A~.
\eqe
The area constraint is not considered further in this paper. In principle, they can be treated in an equivalent manner to volume constraints.

\subsection{Contact constraints}

Contact is characterized by the impenetrability constraint
\eqb{l}
g_\mrn \leq 0~,
\label{e:cc}\eqe
where
\eqb{l}
g_\mrn := (\bx - \bx_\mrp)\cdot\bn_\mrp
\label{e:gn}\eqe
denotes the normal gap between the membrane point $\bx\in\sS$ and the surface $\Gamma$ of a neighboring obstacle. Here, the unit vector $\bn_\mrp$ denotes the surface normal of $\Gamma$ at the point $\bx_\mrp$, which is the solution of the minimum distance problem 
\eqb{l}
\bx_\mrp(\bx) = \big\{\by\big|\ds\min_{\forall\by\in\Gamma}(\bx-\by)~$for$~\bx\in\sS\big\}~.
\eqe
We note that this minimization can cause difficulties for complex surface geometries \citep{wriggers-contact}.
Constraint~(\ref{e:cc}) can be included in the membrane formulation by various methods. 
The simplest of these is the penalty formulation. 
In this case the contact traction $\bt_\mrc$ acting at $\bx\in\sS$ is given by
\eqb{l}
\bt_\mrc = \left\{\begin{array}{ll}
-\epsilon_\mrn\,g_\mrn\,\bn_\mrp~, & g_\mrn<0~, \\
\mathbf{0}~, & g_\mrn\geq 0~,
\end{array}\right.
\label{e:PM}\eqe
where $\epsilon_\mrn$ is the penalty parameter.
The contact forces contribute to virtual work balance~(\ref{e:wf_3}). This can be expressed by including 
\eqb{l}
G_\mrc = -\ds\int_\sS\bw\cdot\bt_\mrc\,\dif a
\label{e:Gc}\eqe
on the right hand side of Eq.~(\ref{e:wf_3}). For two deformable membranes in contact, weak form~(\ref{e:wf_3}) must be satisfied for each membrane, and contribution $G_\mrc$ is added correspondingly to each weak form. To avoid a surface bias it is advantageous to treat both contact pairs equivalently as is done in the \textit{two-half-pass algorithm} of \citet{spbc} . For some problems, like adhesion, the contact constraint is replaced by suitable constitutive contact laws of the form $\bt_\mrc=\bt_\mrc(g_\mrn)$, see \citet{sauer07b}.


\section{Finite element discretization}\label{s:fed}

The governing equations~(\ref{e:wf_3}) and (\ref{e:gv}) are solved by the finite element (FE) method. The initial surface $\sS_0$ is therefore discretized into a set of finite elements $\Omega_0^e$ that are defined by nodal points $\bX_I$ or control points in the case of isogeometric FE. The deforming membrane is then described by the motion of the nodal points $\bX_I\rightarrow\bx_I$, which corresponds to a Lagrangian FE description. The deformed configuration of element $\Omega^e_0$ is denoted $\Omega^e$. Here we consider quadrilateral elements since these can be conveniently related to a master element in the parameter domain $\xi^\alpha\in[-1,1]$.

\subsection{Finite element interpolation}\label{s:fei}

Within elements $\Omega_0^e$ and $\Omega^e$, the geometry is approximated by the nodal interpolations
\eqb{l}
\ds \bX \approx \bX^h=\sum_I N_I\, \bX_I~,
\label{e:Xapprox}\eqe
and
\eqb{l}
\ds \bx \approx \bx^h=\sum_I N_I\, \bx_I~,
\label{e:xapprox}\eqe
where $N_I=N_I(\xi^1,\xi^2)$ denotes the nodal shape function defined on the master element in parameter space. 
The summation is carried out over the $n_\mathrm{ne}$ nodes of the element. 
Here, the following quadrilateral elements are considered: 4-noded linear Lagrange elements, 9-noded quadratic Lagrange elements, quadratic NURBS elements, and T-spline elements. In principle, any other element type can also be considered. For isogeometric elements the shape functions are constructed via the B\'ezier extraction operation \citep{borden11,scott11}. According to Eq.~(\ref{e:a_a}), the tangent vectors are thus approximated by
\eqb{l}
\ds\ba_{\alpha} \approx \sum_I N_{I,\alpha}\, \bx_I~,
\label{e:aapprox}\eqe
where $N_{I,\alpha}=\partial N_I/\partial\xi^\alpha$.
Considering a Bubnov-Galerkin formulation the variation $\bw$ is approximated in the same way as the deformation, i.e.
\eqb{l}
\ds\bw \approx \sum_I N_I\,\bw_I~.
\label{e:wapprox}\eqe
For shorthand notation, we rewrite Eqs.~(\ref{e:Xapprox}), (\ref{e:xapprox}) and (\ref{e:wapprox}) 
\eqb{l}
\bX\approx\mN\mX_e~,\quad
\bx\approx\mN\mx_e~,\quad
\bw\approx\mN\mw_e~,
\eqe
where $\mN:=[N_1\tilde\bone,\,...,\,N_I\tilde\bone,\, ...]$ is a ($3\times3n_\mathrm{ne}$) array with the usual identity tensor $\tilde\bone$ and $\mX_e$, $\mx_e$, and $\mw_e$ are vectors containing the stacked nodal values for the element.\footnote{Non-italic discrete arrays $\mX$, $\mx$, $\mw$ and $\mN$ should not be confused with italic field variables $\bX$, $\bx$, $\bw$ and $\bN$.}
In order to discretize the weak form we need to discretize $w_\alpha$, $w$ and $\bw_{;\alpha}$. We find
\eqb{lllll}
w_{\alpha} \ais \bw\cdot\ba_{\alpha}  \is \mw^T_e\mN^T\ba_{\alpha} = \mw^T_e\mN^T\mN_{,\alpha}\,\mx_e \\[1mm]
w \ais \bw\cdot\bn \is \mw^T_e\mN^T\bn \\[1mm]
\bw_{;\alpha} \ais \mN_{,\alpha}\,\mw_e \back\back
\label{e:wapprox2}\eqe
where $\mN_{,\alpha}:=[N_{1,\alpha}\tilde\bone,\,...\,N_{I,\alpha}\tilde\bone,\,...]$. Note that for a vector like $\bw$, the co-variant derivative $\bw_{;\alpha}$ coincides with the regular partial derivative $\bw_{,\alpha}$. The surface normal $\bn$ is given through definition (\ref{e:n}) and approximation (\ref{e:aapprox}). 
According to (\ref{e:b_ab}), the components of the curvature tensor become
\eqb{l}
b_{\alpha\beta} \approx \bn\cdot\mN_{,\alpha\beta}\,\mx_e~.
\label{e:bapprox}\eqe
With the above expressions, we further find
\eqb{l}
w_{\alpha;\beta} \approx \mw_e^T\Big(\mN^T_{,\beta}\,\mN_{,\alpha} 
+ \mN^T(\bn\otimes\bn)\,\mN_{,\alpha\beta}\Big)\mx_e
\label{e:wapprox3}\eqe
according to eq.~(\ref{e:w_ab}).

\subsection{Discretized weak form}\label{s:dwf}

The above expressions are now used to discretize the membrane weak form of Sec.~\ref{s:wf}. The surface integration 
is carried out over the element domains $\Omega^e$ and then summed over all $n_\mathrm{el}$ FE as
\eqb{l}
G_\mathrm{int} = \ds\sum_{e=1}^{n_\mathrm{el}} G_\mathrm{int}^e~,\quad
G_\mathrm{ext} = \ds\sum_{e=1}^{n_\mathrm{el}} G_\mathrm{ext}^e~.
\eqe
For the internal virtual work of eq.~(\ref{e:Gint2}) we now have
\eqb{l}
G_\mathrm{int}^e = \ds\int_{\Omega^e}\bw_{;\alpha}\cdot\tau^{\alpha\beta}\ba_{\beta}\,\dif A
\approx \mw_e^T\int_{\Omega^e}\mN^T_{,\alpha}\,\tau^{\alpha\beta}\,\mN_{,\beta}\,\dif A\,\mx_e
\eqe
according to approximations (\ref{e:aapprox}) and (\ref{e:wapprox2}). Writing $G^e_\mathrm{int}=\mw_e^T\,\mf^e_\mathrm{int}$, we identify the internal FE force vector
\eqb{l}
\mf_\mathrm{int}^e = \ds\int_{\Omega^e}\mN^T_{,\alpha}\,\tau^{\alpha\beta}\,\mN_{,\beta}\,\dif A\,\mx_e~.
\label{e:fint}\eqe
As noted in Eq.~(\ref{e:Gintext}.1),
the internal virtual work can be split into in-plane and out-of plane contributions. At the element level these are
\eqb{l}
G_\mathrm{inti}^e = \ds\int_{\Omega^e_0}w_{\alpha;\beta}\,\tau^{\alpha\beta}\, \dif A~,\quad
G_\mathrm{into}^e = - \int_{\Omega^e_0} w\,\tau^{\alpha\beta}\,b_{\alpha\beta}\,\dif A~,
\eqe
such that $G^e_\mathrm{int}=G^e_\mathrm{inti}+G^e_\mathrm{into}$. In view of Eqs.~(\ref{e:wapprox2}), (\ref{e:bapprox}) and (\ref{e:wapprox3}), the corresponding force vectors become
\eqb{lll}
\mf_\mathrm{inti}^e \is \ds\int_{\Omega^e_0}\tau^{\alpha\beta}\Big(\mN^T_{,\alpha}\,\mN_{,\beta} 
+ \mN^T(\bn\otimes\bn)\,\mN_{,\alpha\beta}\Big)\, \dif A\,\mx_e~,\\[4mm]
\mf_\mathrm{into}^e \is -\ds\int_{\Omega^e_0} \tau^{\alpha\beta}\,\mN^T(\bn\otimes\bn)\,\mN_{,\alpha\beta}\,\dif A\,\mx_e~.
\label{e:fintio}\eqe
For $G_\mathrm{ext}$ we consider external loading of the form $\bff = \bff_0/J + p\,\bn$, where $\bff_0$ and $p$ are given loading parameters. The first corresponds to a dead force per reference area, the second to a live pressure. According to Eq.~(\ref{e:Gintext}.2) we then have
\eqb{l}
G_\mathrm{ext}^e = \ds\int_{\Omega^e_0} \bw\cdot\bff_0\,\dif A
+ \int_{\partial_t\Omega^e}\bw\cdot\bar\bt\,\dif s
+ \int_{\Omega^e} w\,p\,\dif a~,
\eqe
which yields the external force vector
\eqb{l}
\mf_\mathrm{ext}^e = \ds\int_{\Omega^e_0} \mN^T\,\bff_0\,\dif A
+ \int_{\partial_t\Omega^e}\mN^T\,\bar\bt\,\dif s
+ \int_{\Omega^e} \mN^T\,p\,\bn\,\dif a~.
\label{e:fext}\eqe
The original weak form~(\ref{e:wf_3}) now yields the  descretized version
\eqb{l}
\mw^T\big[\mf_\mathrm{int}-\mf_\mathrm{ext}\big] = \mathbf{0}~,
\label{e:dwf}\eqe
where $\mf_\mathrm{int}$ and $\mf_\mathrm{ext}$ are obtained from the assembly of the corresponding elemental force vectors, and $\mw$ is the kinematically admissible set of all nodal variations. These are zero for the nodes on the Dirichlet boundary $\partial_u\sS$. For the remaining nodes, Eq.~(\ref{e:dwf}) implies
\eqb{l}
\mf := \mf_\mathrm{int}-\mf_\mathrm{ext} = \mathbf{0}~,
\label{e:dequilib}\eqe
which is the discretized equilibrium equation that needs to be solved for the unknown nodal positions $\mx$; see Sec.~\ref{s:sol}. 

We note, that in this formulation no mapping of derivatives between master and current configuration is required. Also no introduction of local, cartesian bases are needed. The formulation thus is straight forward and efficient to implement.

\subsection{Contact contributions}\label{s:dc}

The proposed membrane model can be easily extended to include contact, provided a 3D contact algorithm is available.
The contact contribution~(\ref{e:Gc}) simply yields the force vector
\eqb{l}
\mf^e_{\mrc} = -\ds\int_{\Omega^e_k}\mN^T\,\bt_\mrc\,\dif a~,
\label{e:fc}\eqe
that needs to be included in Eq.~(\ref{e:dequilib}).
%
For details on the the FE implementation of Eq.~(\ref{e:fc}) we refer to \citet{spbc}.

\subsection{Discretized volume constraint}

The volume, enclosed by the discretized membrane surface, is obtained as
\eqb{l}
V = \ds\frac{1}{3}\sum_{e=1}^{n_\mathrm{el}}\int_{\Omega^e}\bn^T\mN\,\dif a\,\mx_e
\eqe
according to Eq.~(\ref{e:V}). For the volume constraint, $g_\mrv=V-V_0=0$, $V_0$ can be considered as an externally prescribed volume, e.g. during inflation, or as the initial value of $V$.

\subsection{Solution method}\label{s:sol}

The volume constraint is included in the formulation by the Lagrange multiplier method. The Lagrange multiplier associated with the constraint is the pressure, $p$, acting on membrane. Combining (\ref{e:gv}) with (\ref{e:dequilib}) leads to the system 
\eqb{lll}
\mf(\mx,p) \is \mathbf{0}~, \\[1mm]
g_\mrv(\mx) \is 0~,
\eqe
that needs to be solved for the unknown nodal position $\mx$ and pressure $p$. Due to the nonlinearities of the model, this is
solved with Newton's method. Therefore, the linearization of $\mf$ and $g_\mrv$ w.r.t. $\mx$ and $p$ are needed. This is discussed in Appendix~\ref{s:LFE}.

\subsection{Hydrostatic pressure}

In some applications, the pressure $p$ may vary locally. In static examples this is typically due to gravity. An example is the hydrostatic pressure distribution in a water-filled membrane. In this case, we have
\eqb{l}
p = p_\mrv + p_\mrh~,
\eqe
where $p_\mrh$ is the hydrostatic, height dependent, pressure and $p_\mrv$ is the pressure associated with the volume constraint.
The former is simply written as
\eqb{l}
p_\mrh = -\rho\,\bg\cdot\bx
\label{e:ph}\eqe
where $\rho$ is the density of the pressure causing medium, and $\bg$ is the gravity vector.\footnote{typically $\bg=-[0,~0,~g]^T$, where $g$ is the gravity constant} The value of $p_\mrv$ is then the (constant) datum pressure at the origin.

\subsection{Numerical quadrature}

In parameter space, each element is defined on the master domain $\xi^\alpha\in[-1,1]$, $\alpha=1,2$. The integrals from above are mapped to the master domain using transformations~(\ref{e:da}). Integration is then carried out with standard Gaussian quadrature on the master domain.

\subsection{Monitoring compression}

Membranes do not support in-plane compression. The absence of physical bending stiffness leads to buckling of the structure,  known as wrinkling in the case of membranes. To avoid membrane compression in our formulation during computations, we simply monitor the minimum principal stress
\eqb{l}
\sig_\mathrm{min} = \ds \frac{I_1}{2} - \sqrt{\frac{I_1^2}{4} - I_2}~,
\eqe
where $I_1=\mathrm{tr}\,\bsig = \sig_\alpha^\alpha$ and $I_2= \det\bsig = \det\sig^\alpha_\beta$
are the two invariants of the surface stress tensor. We note that $\sig_\mathrm{min}$ does not imply the automatic failure of the discretize membrane structure as some numerical bending stiffness may be present.
More involved wrinkling criteria can be found in the literature, see \citet{lu01} and \citet{youn06}.


\section{Numerical examples}\label{s:ne}

The proposed membrane model is illustrated by several examples, considering both solid and liquid membranes under dead, pressure, and volume loading.
Standard linear and quadratic Lagrange finite elements as well as quadratic NURBS and cubic T-spline finite elements, providing $C^1$- and $C^2$-continuous surface descriptions, respectively, are used.

\subsection{Inflation of a spherical balloon}\label{s:balloon}

\begin{figure}[h]
\begin{center} \unitlength1cm
\includegraphics[height=60mm]{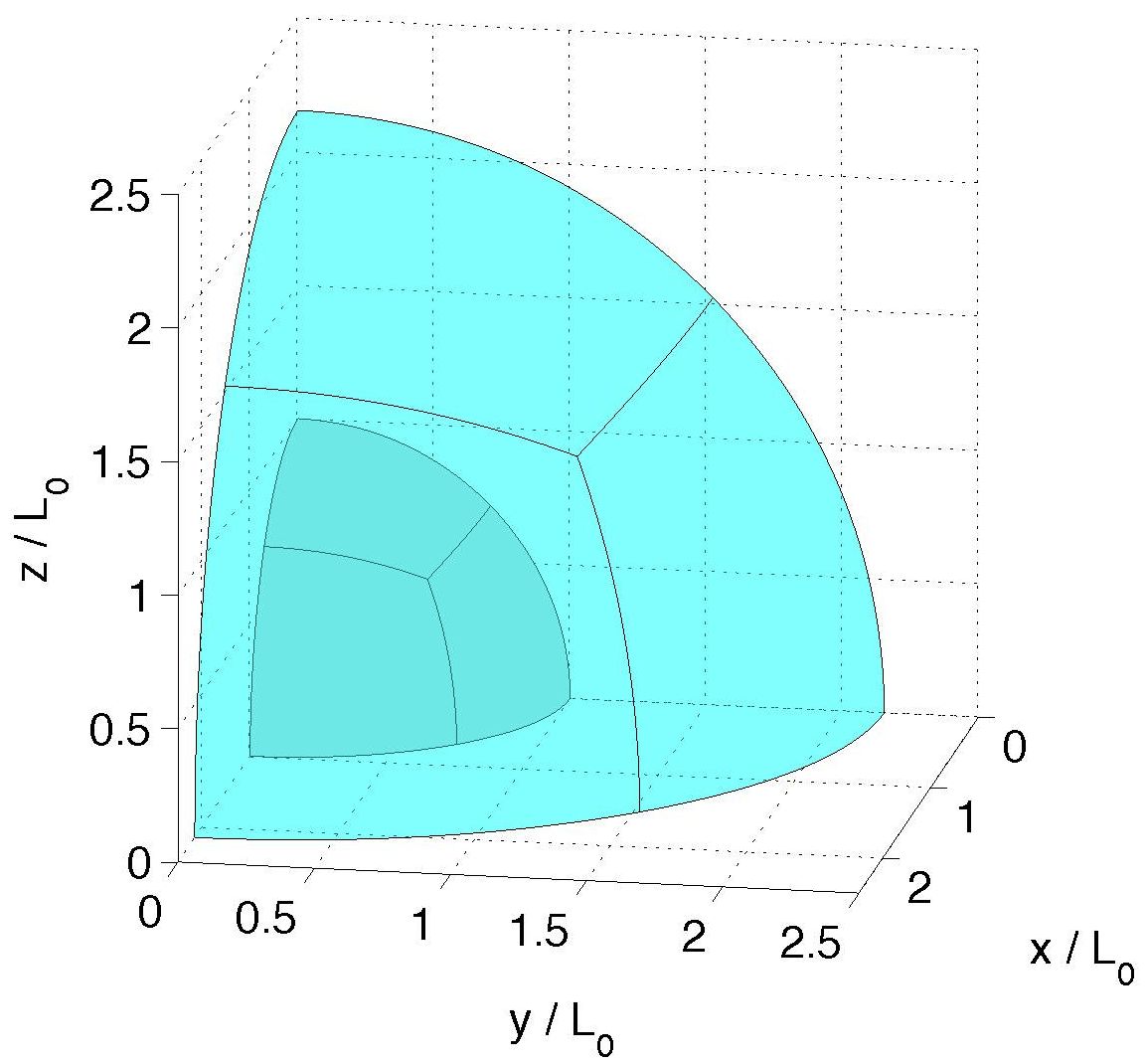}
\includegraphics[height=65mm]{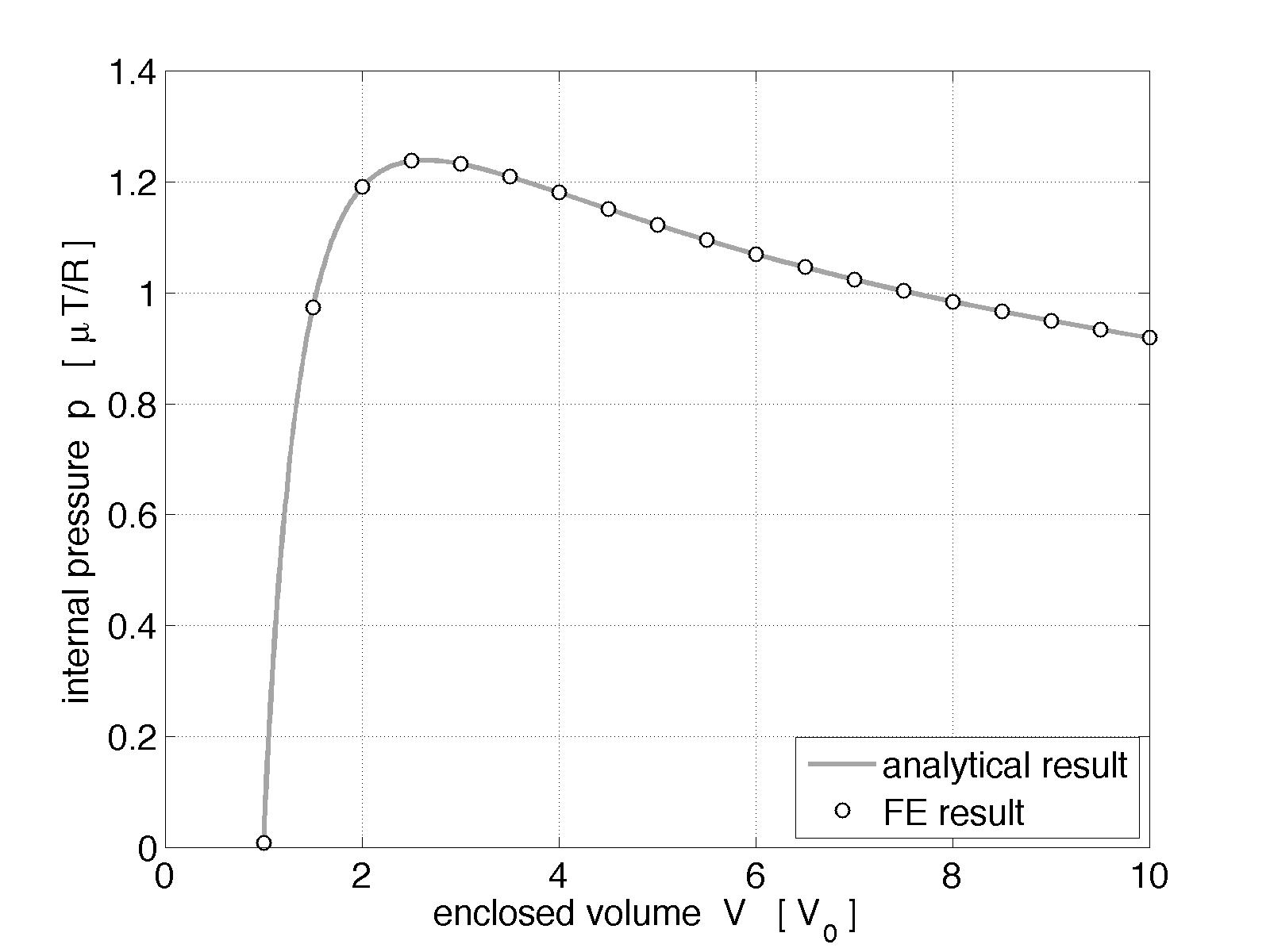}
\put(-15.2,0){a.}
\put(-8.3,0){b.}
\caption{Inflated balloon: (a) initial and current configuration (for $V=10\,V_0$); (b) pressure-volume relation for $V\in[1~10]V_0$ (FE result for 3 quadratic FE).}
\label{f:balloon_inflation:load}
\end{center}
\end{figure}

\begin{figure}[h]
\begin{center} \unitlength1cm
\includegraphics[height=55mm]{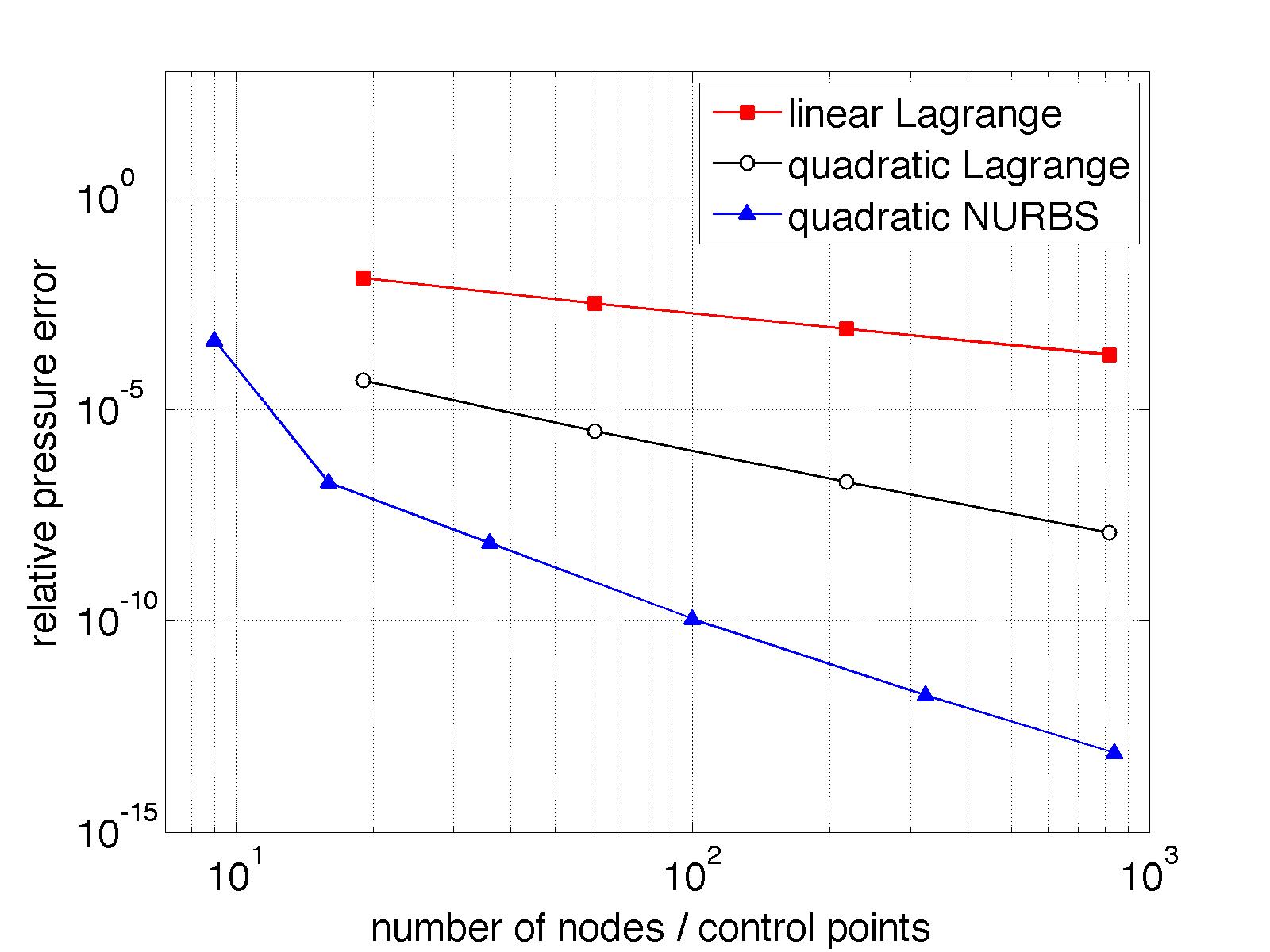}
\includegraphics[height=55mm]{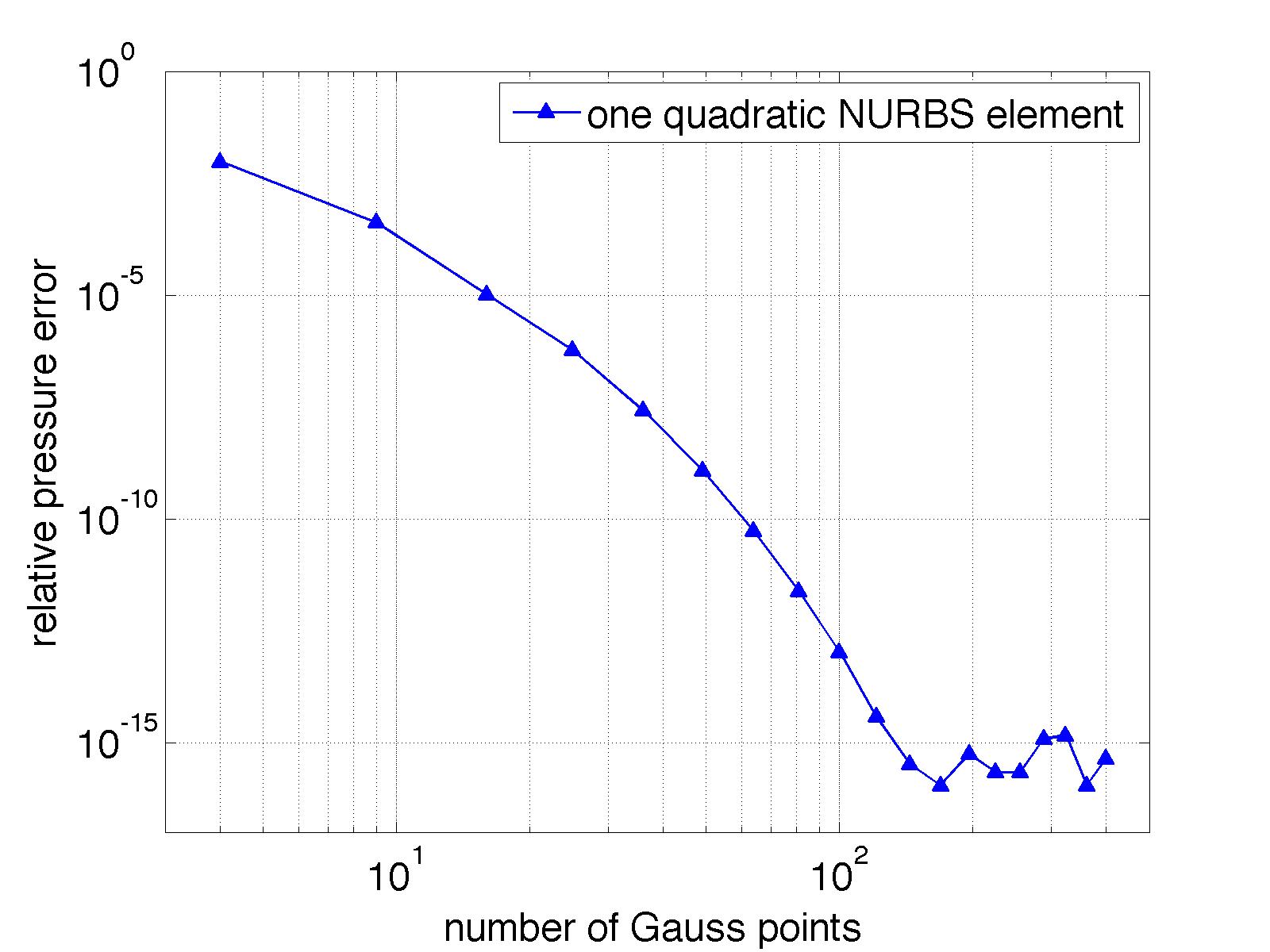}
\put(-15.2,0){a.}
\put(-7.6,0){b.}
\caption{Inflated balloon: convergence of the pressure at $V=10\,V_0$: (a) convergence with mesh size; (b) NURBS convergence with quadrature accuracy.}
\label{f:balloon_inflation:conv}
\end{center}
\end{figure}

We first consider the inflation of a spherical rubber balloon and use it for validation, 
since an analytical solution exists for this problem.
The rubber behavior is described by the incompressible Neo-Hookean material model~(\ref{e:incompNH}).
The finite element model of the balloon, modeled as $1/8$th of a sphere, is shown in Fig.~\ref{f:balloon_inflation:load}a.
Appropriate boundary conditions are provided to maintain the symmetry of the inflating structure. 
The relation between current and initial radius is denoted $r=\lambda R$.
The circumference of the balloon, proportional to $r$, is thus stretched by $\lambda$ such that the surface deformation gradient is $\bF=\lambda\bone$ and the area change is given by $J=\lambda^2$. Due to incompressibility this results in the thickness change $t=T/J$.
According to Eq.~(\ref{e:incompMNH}), the in-plane normal stress within the balloon thus is $\tilde\sig=\sig/t = \mu T(1-\lambda^6)/t$, which is equal to the well-known formula $\tilde\sig=pr/2/t$.
We thus obtain the pressure-stretch relation
\eqb{l}
\ds\frac{pR}{\mu T} = 2\bigg(\frac{1}{\lambda}-\frac{1}{\lambda^7}\bigg)~,
\eqe
or, equivalently, the pressure-volume relation
\eqb{l}
\ds\frac{pR}{\mu T} = 2\bigg(\Big(\frac{V_0}{V}\Big)^{\frac{1}{3}}-\Big(\frac{V_0}{V}\Big)^{\frac{7}{3}}\bigg)~,
\eqe
where $V_0=4\pi/3\,R^3$ is the initial balloon volume. The $p-V$ relation is shown in Fig.~\ref{f:balloon_inflation:load}b.
The pressure increases quickly, peaks and then decreases gradually. This behavior is typical for the inflation of rubber membranes. The FE computation of such problems should therefore be carried out by prescribed volume loading instead of prescribed pressure loading. The proposed FE formulation can capture the analytical behavior very nicely. This is shown by the convergence plot of Fig.~\ref{f:balloon_inflation:conv}a.
Here, the number of Gaussian quadrature points per elements are 2$\times$2 for linear Lagrange and 3$\times$3 for quadratic Lagrange and NURBS elements. Since NURBS elements describe the spherical geometry exactly, they can solve the problem exactly with only one element - provided sufficiently many quadrature points are used. This is shown in Fig.~\ref{f:balloon_inflation:conv}b. The results shown here validate the proposed membrane formulation.

\begin{figure}[h]
\begin{center} \unitlength1cm
\begin{picture}(0,3.9)
\put(2.3,-.5){\includegraphics[height=43mm]{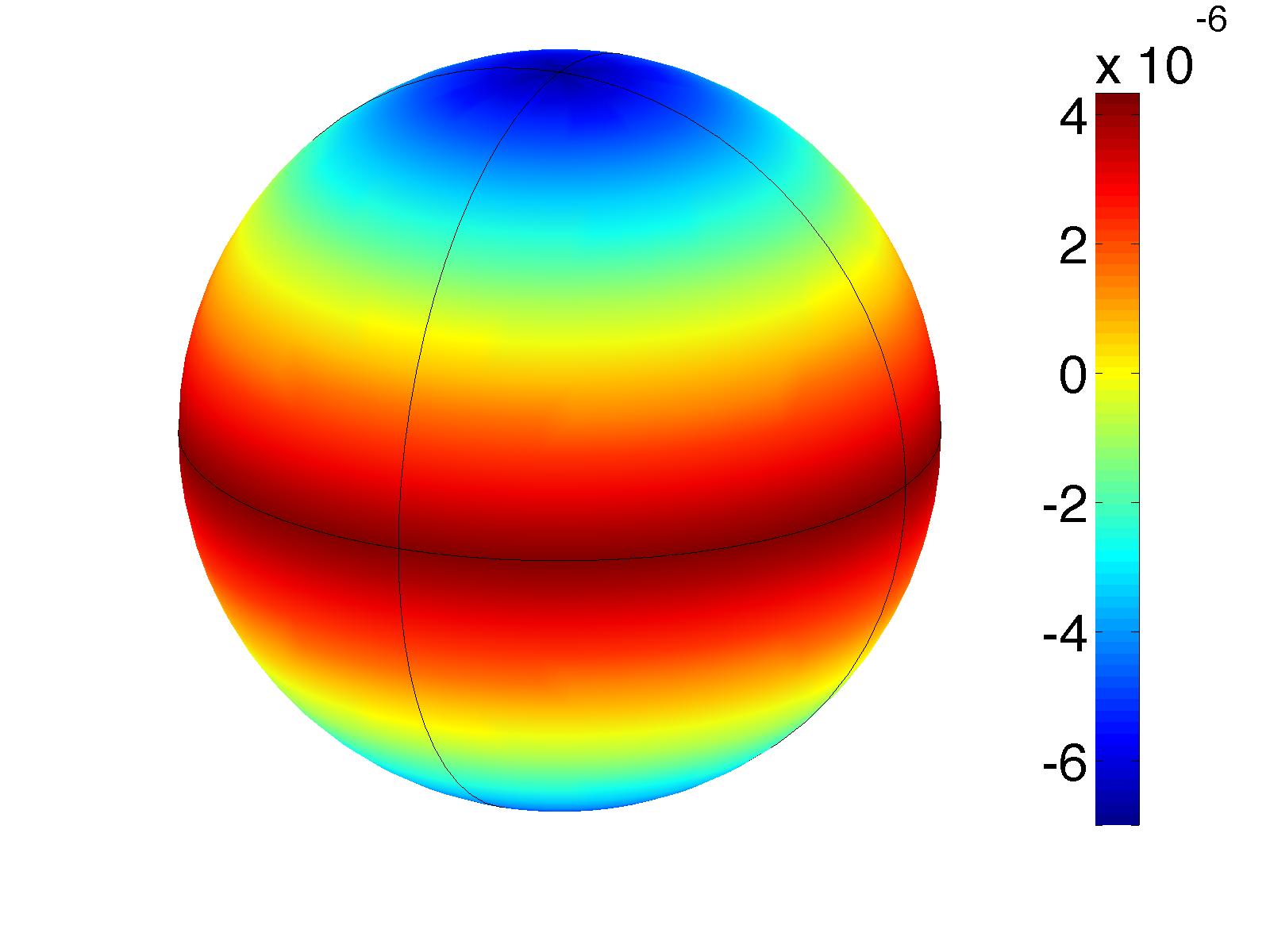}}
\put(-3.2,-.5){\includegraphics[height=43mm]{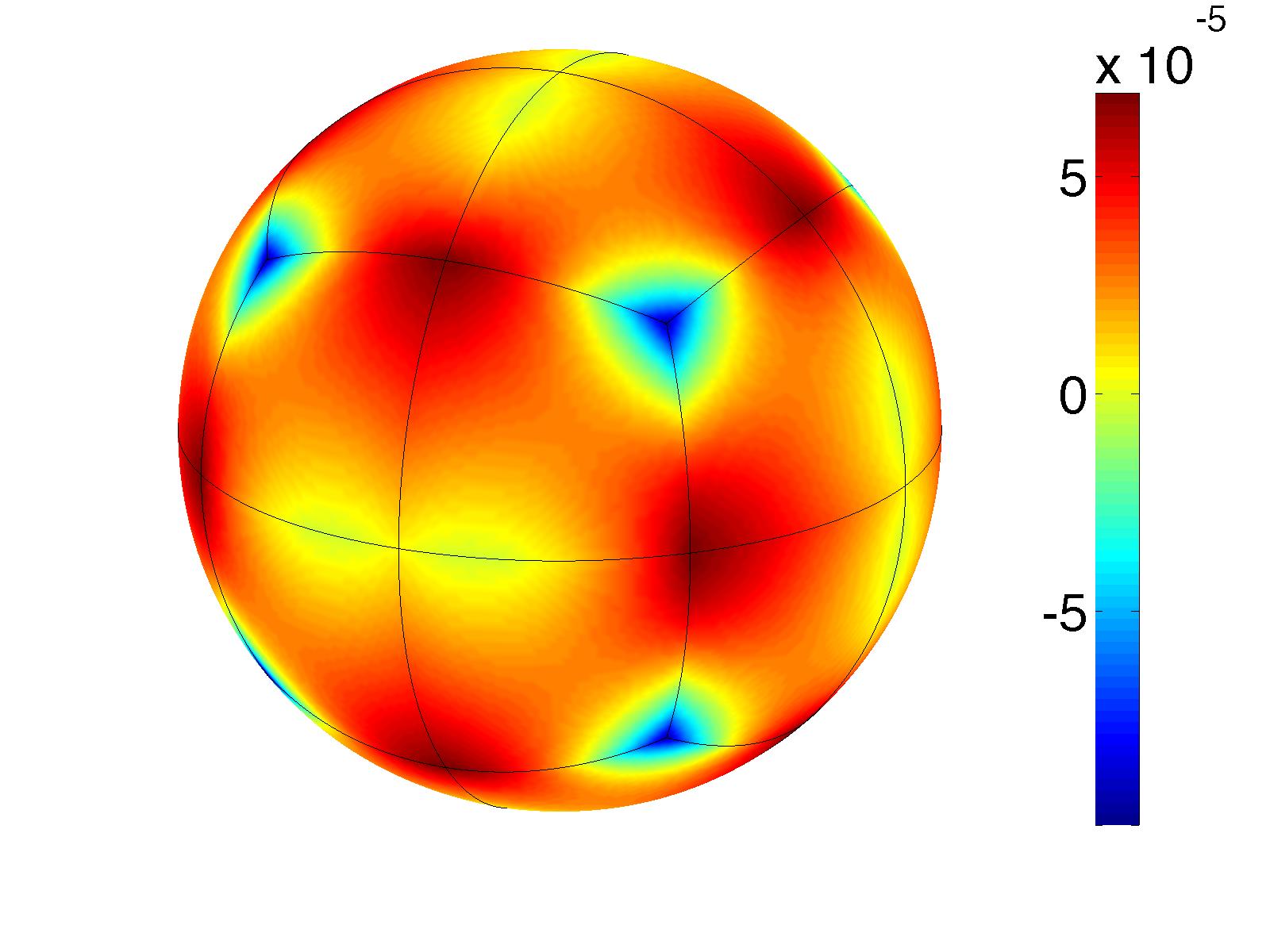}}
\put(-8.7,-.5){\includegraphics[height=43mm]{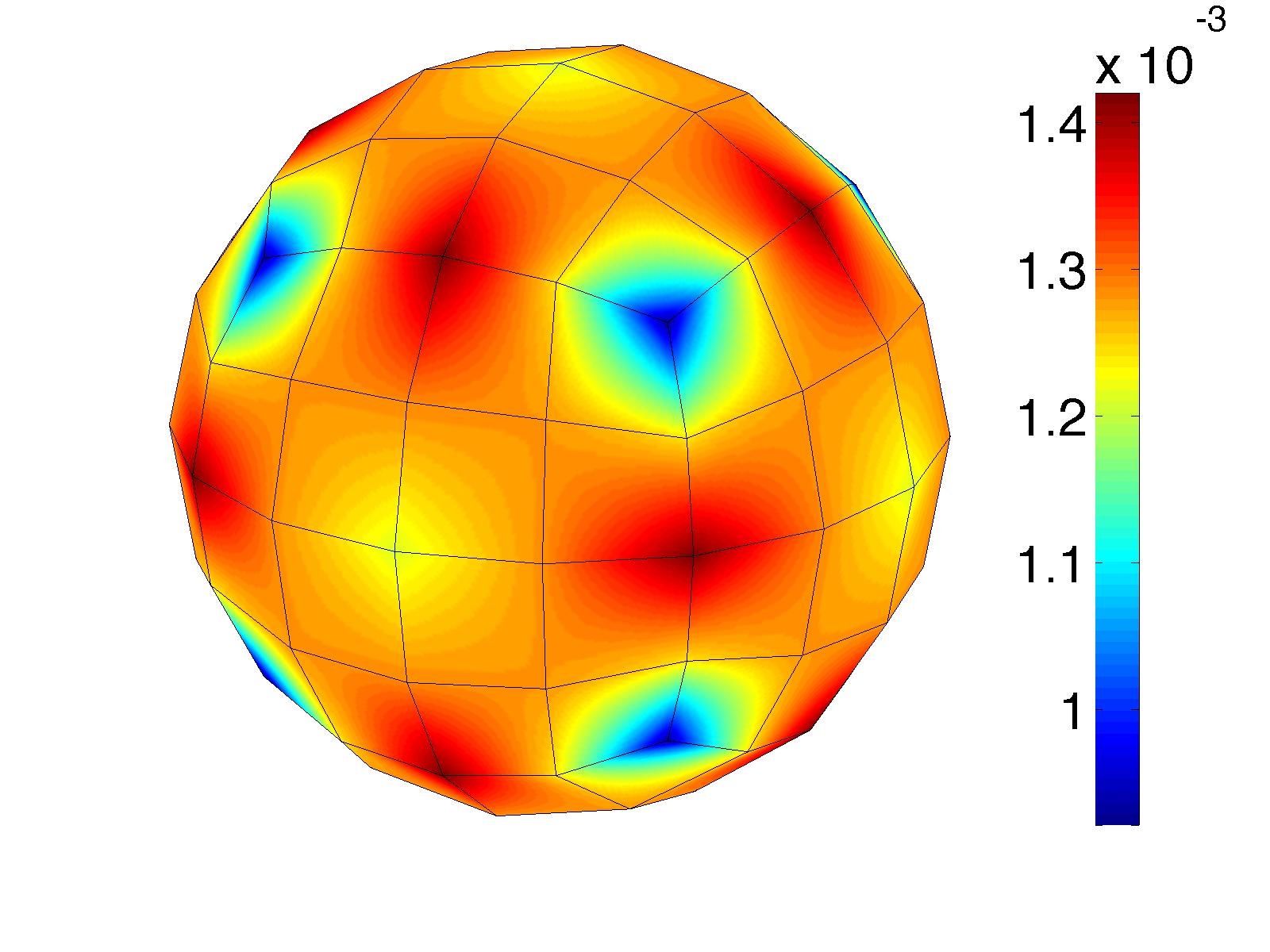}}
\put(-7.9,0){a.}
\put(-2.4,0){b.}
\put(3.1,0){c.}
\end{picture}
\caption{Inflated balloon: Error in the in-plane stress $\sigma=pr/2$ for: (a) 12$\times$8 linear FE, (b) 3$\times$8 quadratic FE, (c) 1$\times$8 NURBS FE.}
\label{f:balloon_inflation:mesh}
\end{center}
\end{figure}
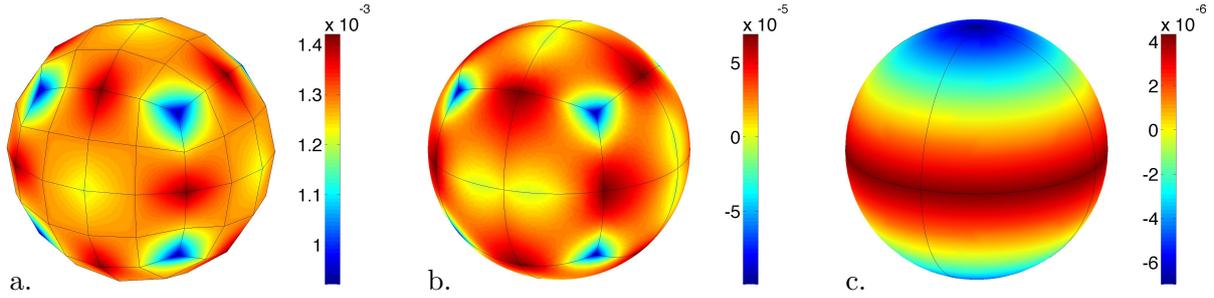

Fig~\ref{f:balloon_inflation:mesh} shows the deformed FE meshes and the error in the membrane stress $\sig$ for the three different element types considered here.
As is seen, the error is smallest for NURBS FE.

\subsection{Inflation of a square sheet}\label{s:sheet}

\begin{figure}[h]
\begin{center} \unitlength1cm
\includegraphics[height=49mm]{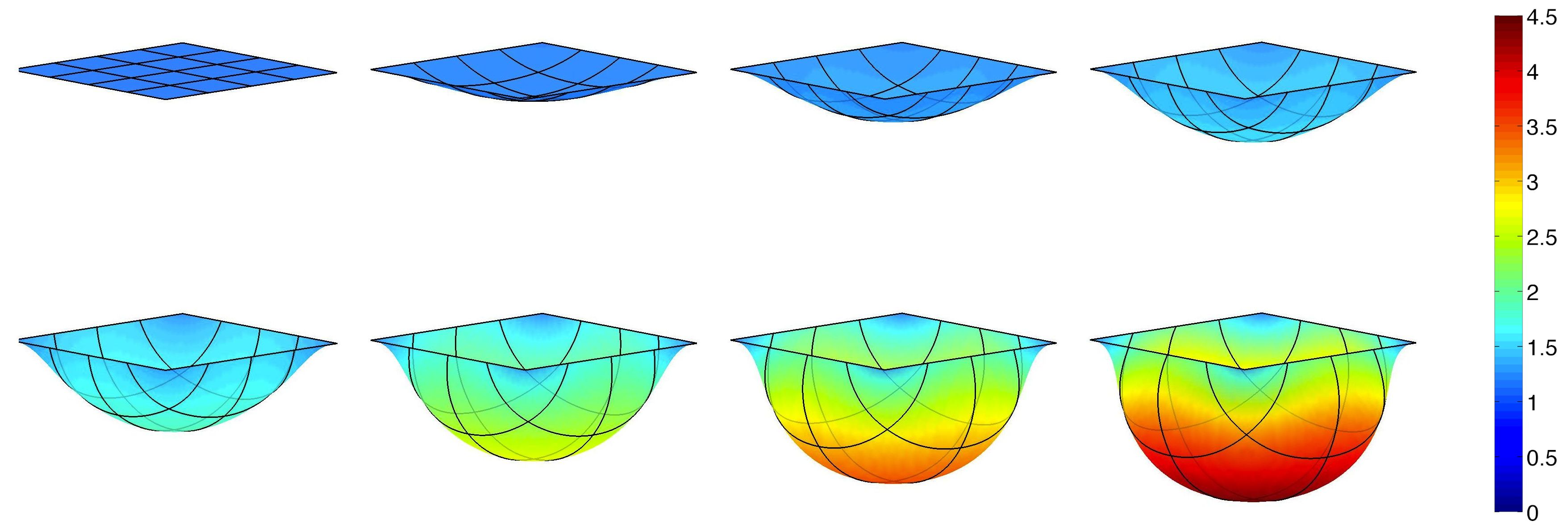}
\caption{Inflated square sheet: configurations for $V=\{0,\,1,\,2,\,3,\,4,\,6,\,8,\,10\}V_0$, where $V_0=4L_0^3$. The coloring shows the area stretch $J$ (which is identical to the thickness decrease).}
\label{f:sheet_inflation:defo}
\end{center}
\end{figure}

As a second example we consider a square membrane sheet with dimension $4\,L_0\times4\,L_0$, apply an isotropic pre-stretch of $\lambda_0=1.05$ to provide initial out-of-plane stiffness, and then inflate the structure by a prescribed volume, as is shown in Fig.~\ref{f:sheet_inflation:defo}.

\begin{figure}[h]
\begin{center} \unitlength1cm
\includegraphics[height=58mm]{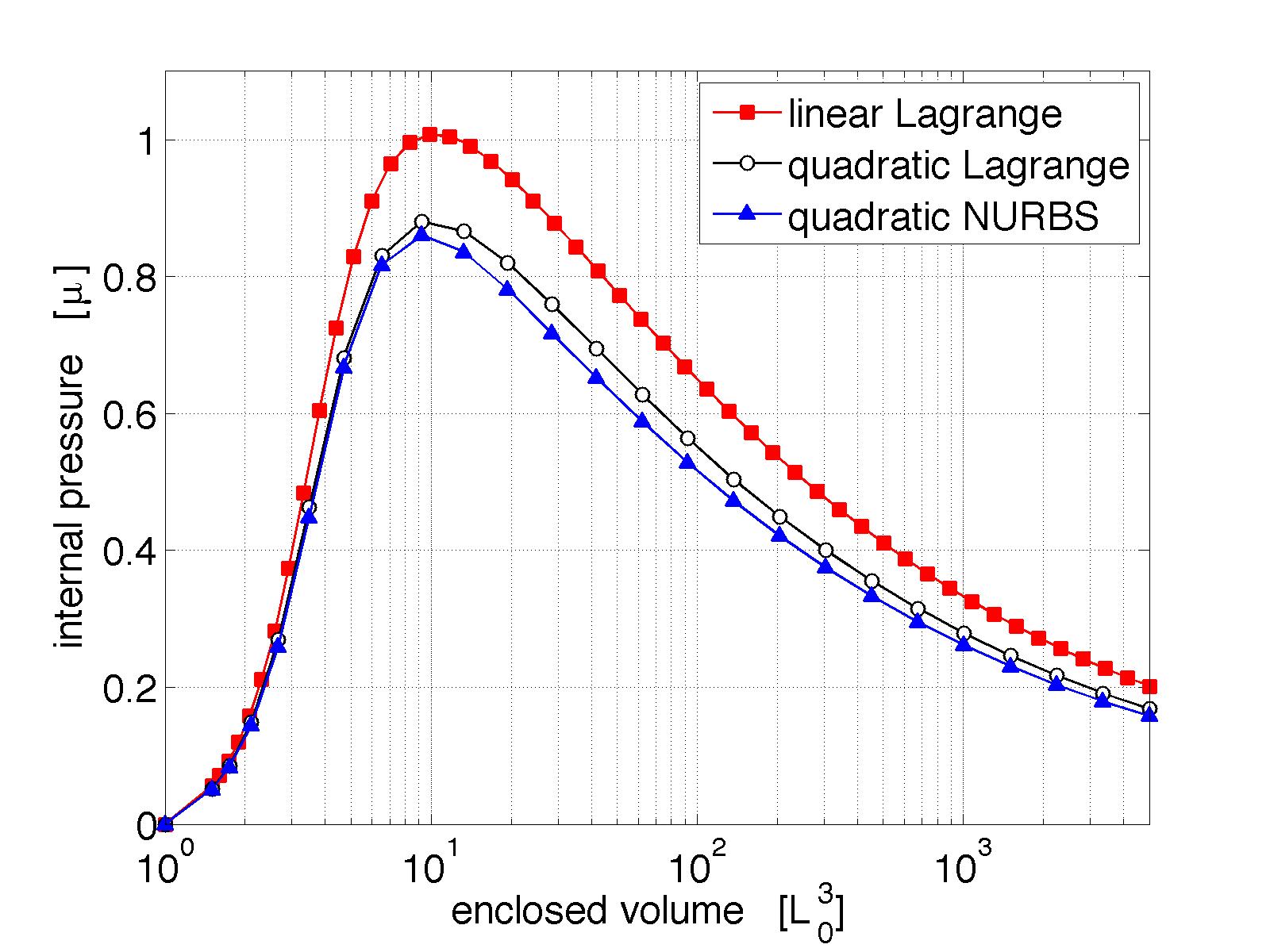}
\includegraphics[height=58mm]{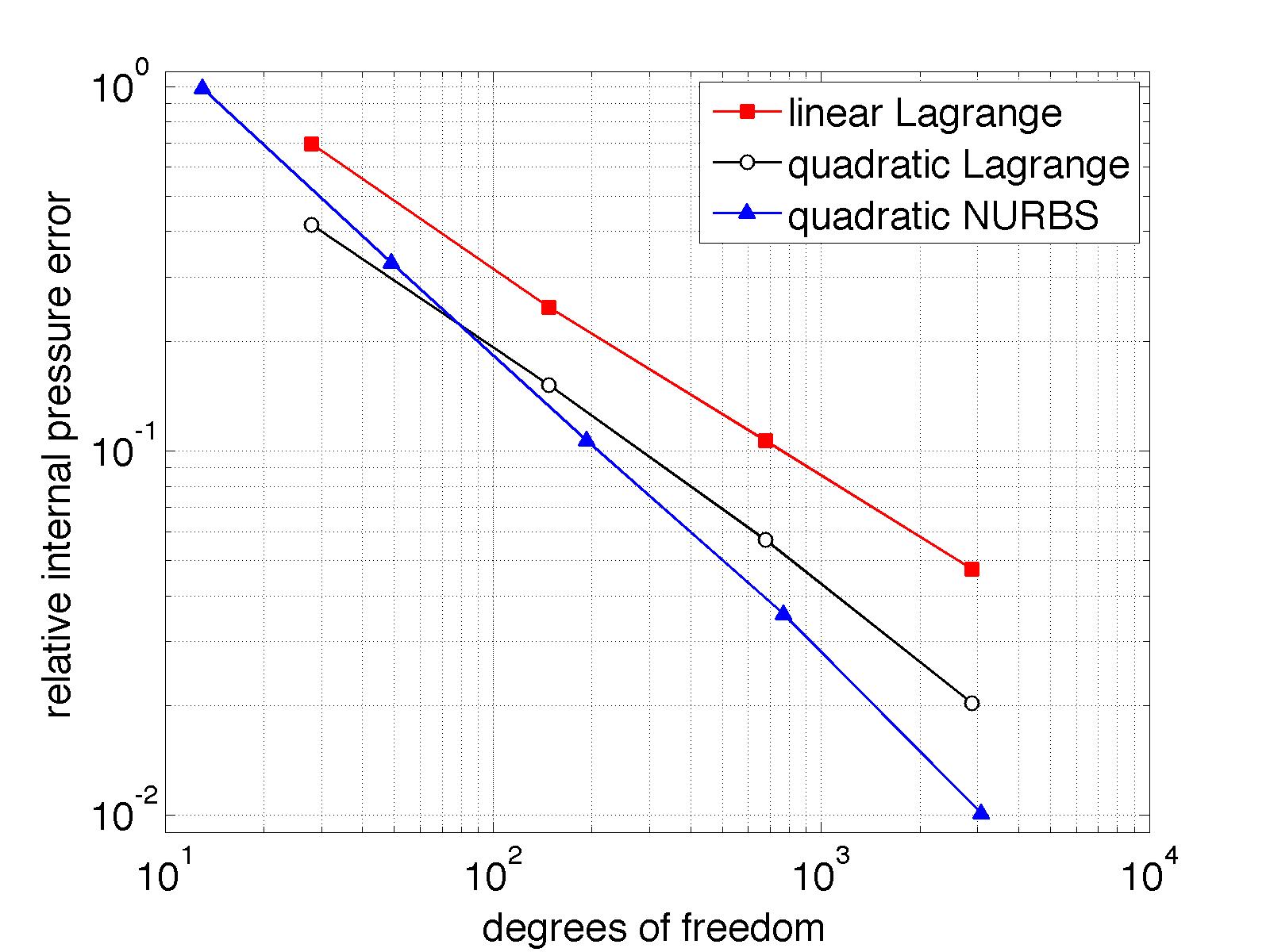}
\put(-15.5,0){a.}
\put(-7.75,0){b.}
\caption{Inflated square sheet: (a) pressure-volume relation; (b) pressure convergence at $V=5000V_0$ (compared to a quadratic NURBS mesh with 23233 dofs).}
\label{f:sheet_inflation:load}
\end{center}
\end{figure}

Fig.~\ref{f:sheet_inflation:load}a shows the pressure-volume relation for the three considered elements.
The accuracy is highest for NURBS elements and lowest for linear elements. This is seen by the convergence behavior of the different element types, shown in Fig.~\ref{f:sheet_inflation:load}b.

\begin{figure}[h]
\begin{center} \unitlength1cm
\begin{picture}(0,5)
\put(-7.8,-.1){\includegraphics[height=37mm]{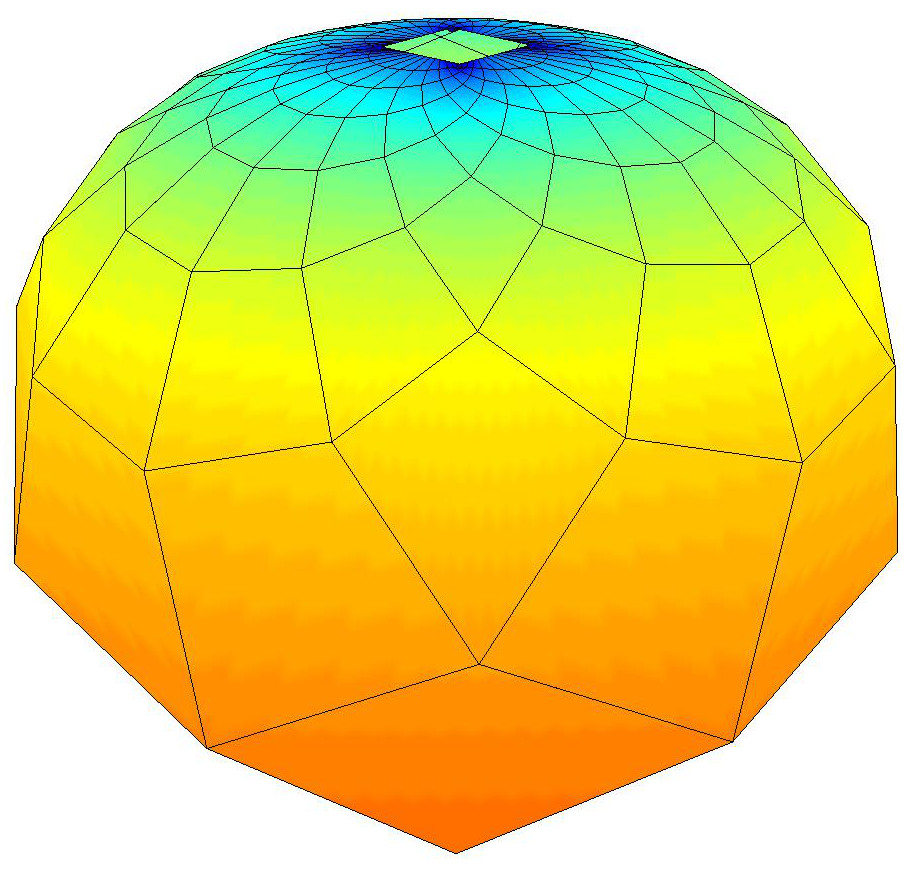}}
\put(-2.7,-.1){\includegraphics[height=37mm]{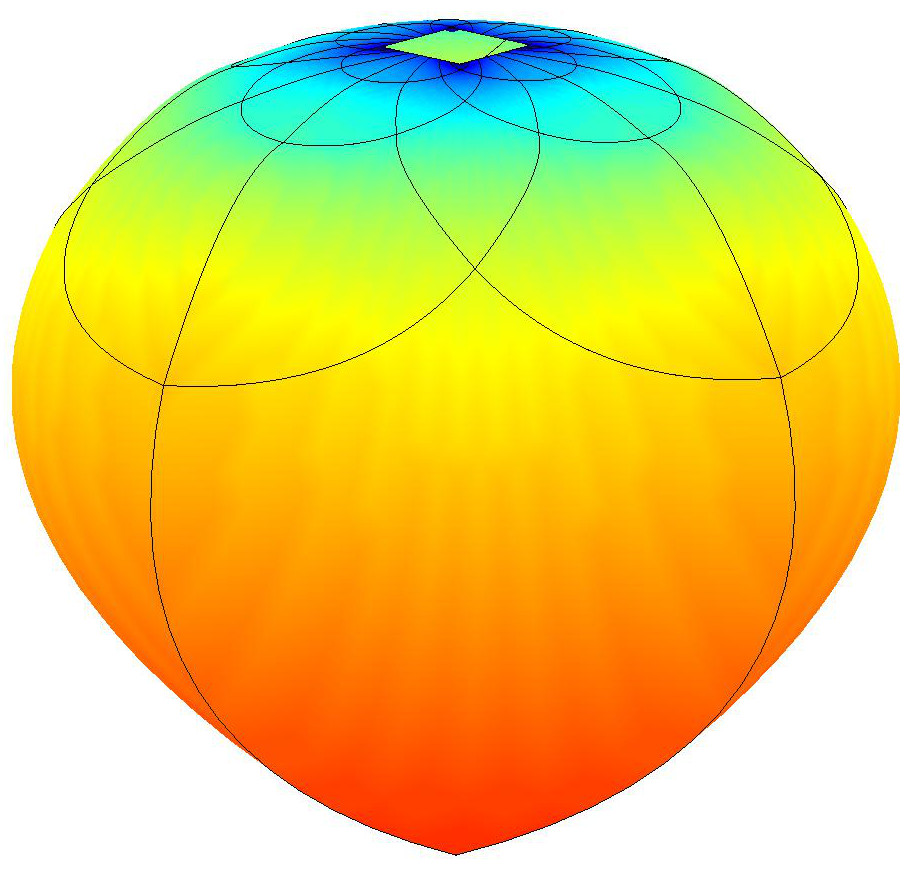}}
\put(2.2,-.1){\includegraphics[height=37mm]{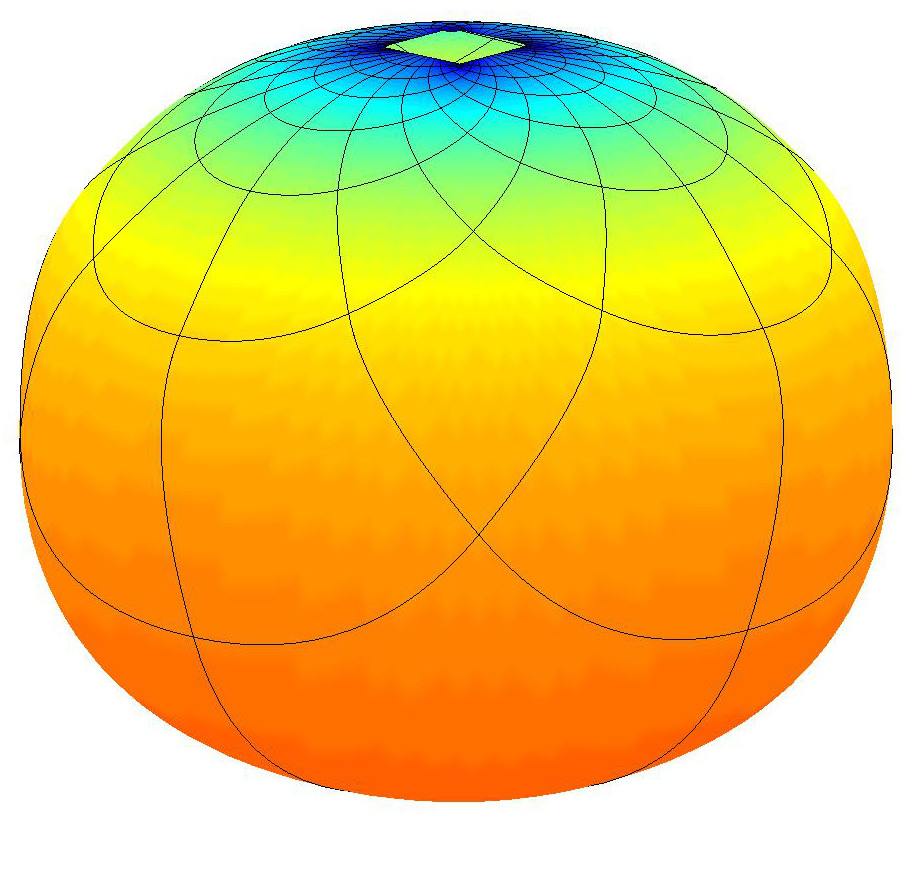}}
\put(7,-.3){\includegraphics[height=40mm]{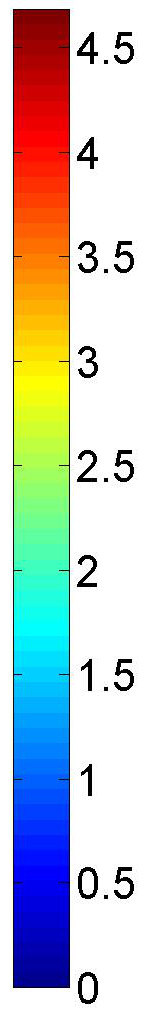}}
\put(-7.9,0){a.}
\put(-2.8,0){b.}
\put(2.2,0){c.}
\end{picture}
\caption{Inflated square sheet: deformation at $V=5000\,V_0$ for (a) $8\times8$ linear elements, (b) $4\times4$ quadratic elements, and (c) $8\times8$ NURBS elements. The color shows the area stretch $J$ displayed as $\log_{10}J$.} 
\label{f:sheet_inflation:elem}
\end{center}
\end{figure}
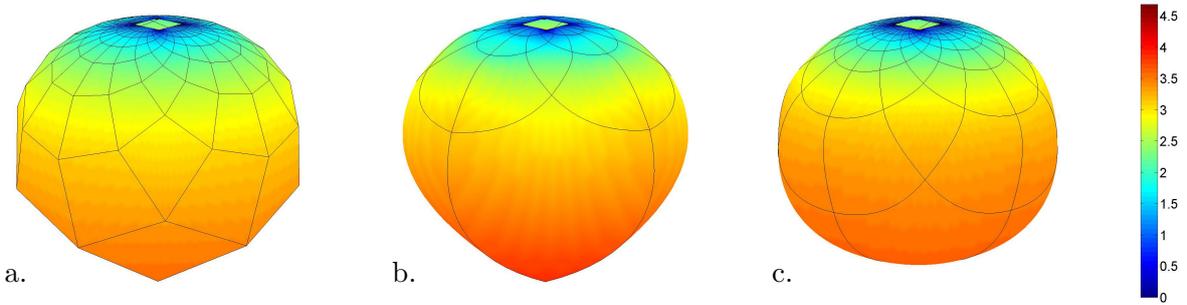

\begin{figure}[h]
\begin{center} \unitlength1cm
\begin{picture}(0,5)
\put(-7.0,-.1){\includegraphics[height=37mm]{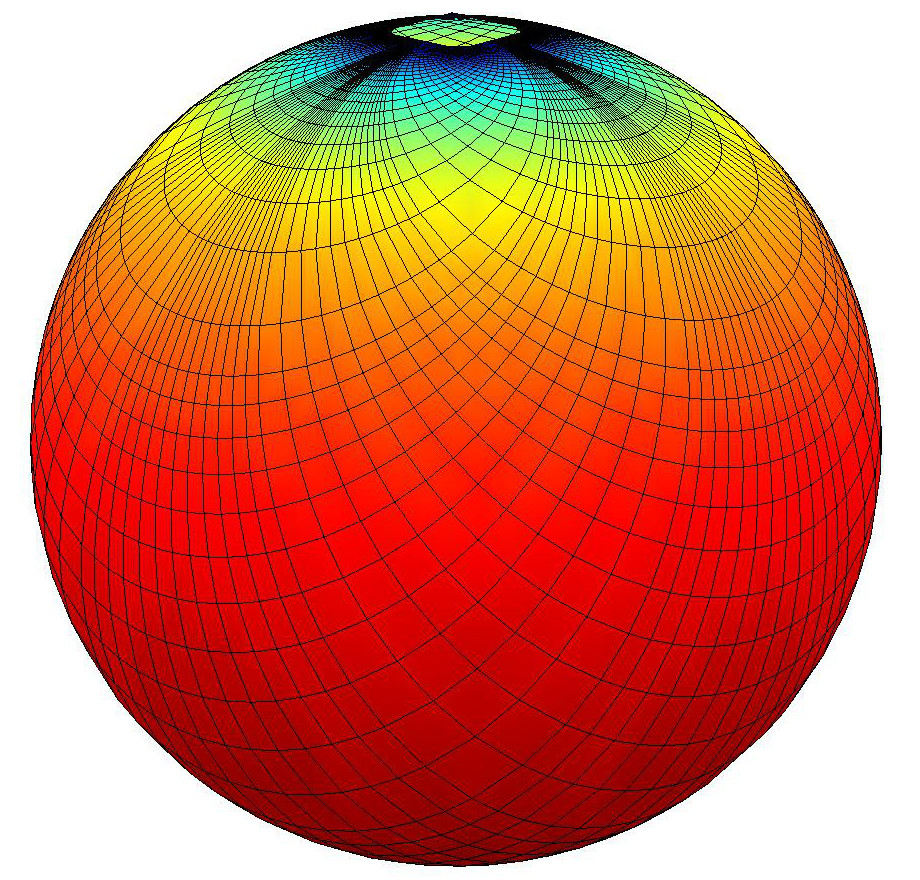}}
\put(-0.5,-.1){\includegraphics[height=50mm]{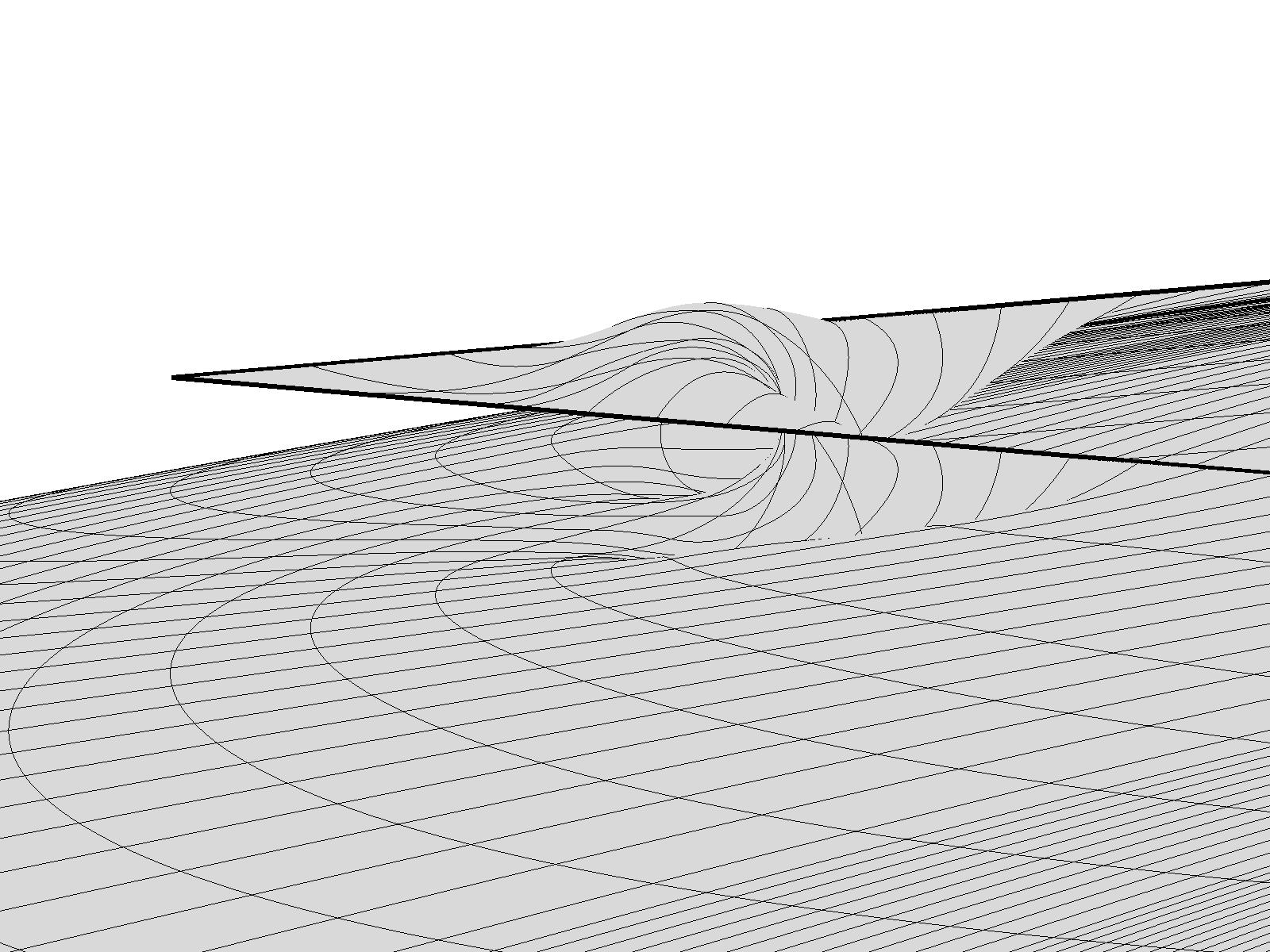}}
\put(-2.4,-.2){\includegraphics[height=40mm]{colorbar_crop.jpg}}
\put(-7.0,0){a.}
\put(-1.2,0){b.}
\end{picture}
\caption{Inflated square sheet: deformation at $V=5000\,V_0$ for $88\times 88$ NURBS elements: (a) overall deformation, (b) deformation at the corner.}
\label{f:sheet_inflation:elem2}
\end{center}
\end{figure}

Fig.~\ref{f:sheet_inflation:elem} shows the deformed sheet for a prescribed volume of $V=5000\,V_0$ for the three element types.
As seen, all element types can accommodate enormous deformations, even for relatively coarse meshes. The comparison with the fine NURBS mesh in Fig.~\ref{f:sheet_inflation:elem2}a shows that there are still considerable inaccuracies present in all three formulations. The NURBS result is fully $C^1$-continuous. In the example, particularly large deformation occur at the bottom and in the corners of the sheet, as is seen in the close-up of Fig.~\ref{f:sheet_inflation:elem2}b.
The deformation in the corner shown a tendency towards wrinkling. We observed that a further mesh refinement led to non-convergent Newton behavior, indicating instabilities. A computational scheme for wrinkling is required to handle this case.  

\subsection{Contact between balloon and cushion}\label{s:cushion}

\begin{figure}[h]
\begin{center} \unitlength1cm
\includegraphics[height=80mm]{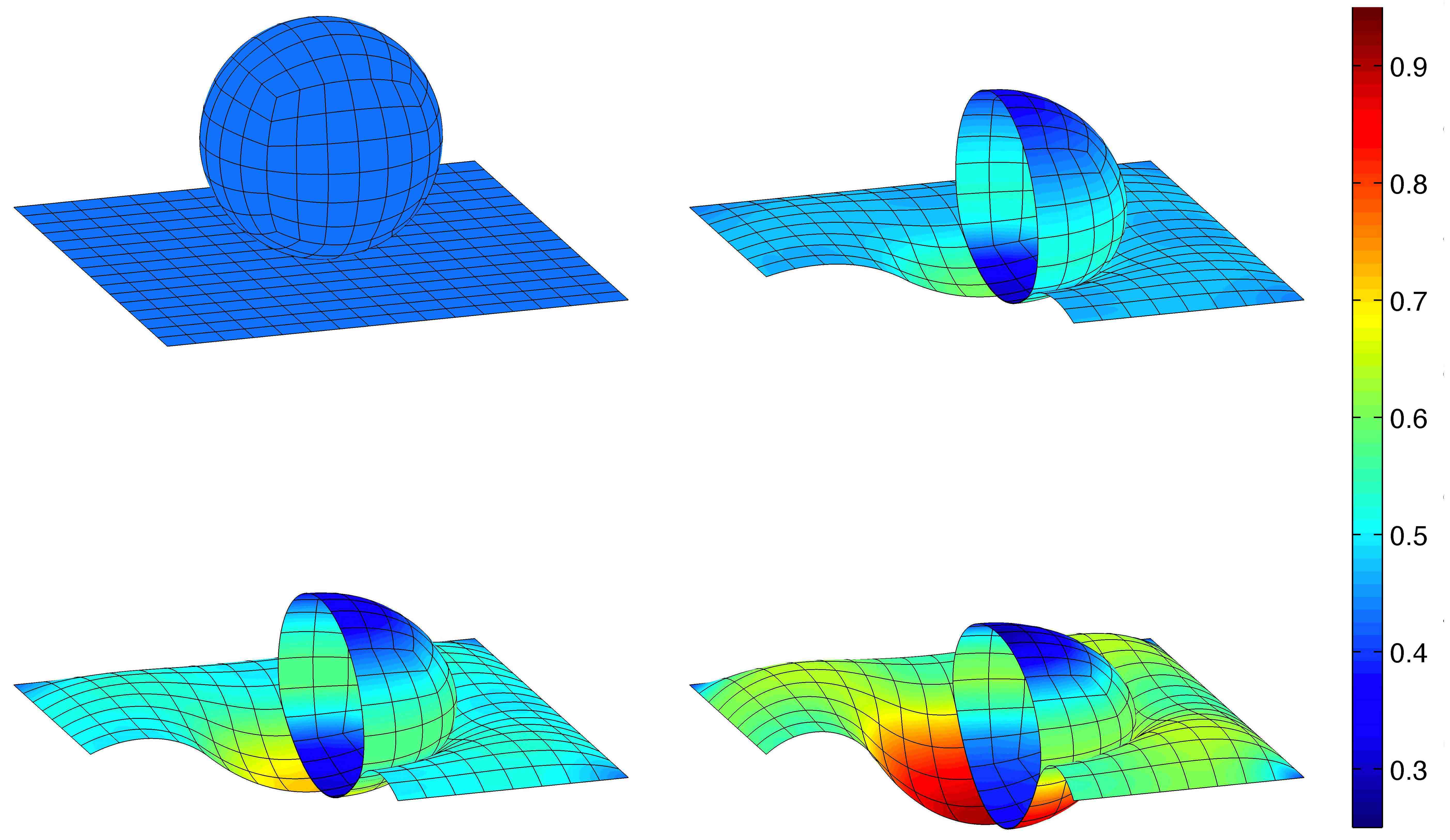}
\caption{Cushion contact: configurations at $\rho g=\{0,\,0.2,\,0.4,\,0.8\}\mu/R$. The coloring shows the stress invariant $I_1=\mathrm{tr}\,\bsig=\sig^\alpha_\alpha$.}
\label{f:cushion_contact:defo}
\end{center}
\end{figure}

The third example considers a spherical, water-filled balloon in contact with a cushion. The balloon is loaded by hydrostatic pressure loading.
The cushion is modeled by a square 
sheet that is fixed along the boundary and supported by internal pressure arising from constraining the volume beneath the sheet. The initial size of the sheet is $2R\times2R$, where $R$ is the undeformed radius of the balloon. Both, balloon and sheet are modeled by material law~(\ref{e:incompNH}) considering equal $\mu$. They are both pre-stretched isotropically by $\lambda_0 = 1.1$, i.e. the constrained balloon volume is $V_0 = 3\pi\,(\lambda R)^3/4$. 
The problem is computed by gradually increasing the gravity level, $g$, pulling on the water inside the balloon. Quadratic finite elements are used. Contact is modeled by the two-half-pass contact algorithm \citep{spbc} considering the augmented Lagrange multiplier method. In principle, any 3D contact algorithm can be applied straight forwardly to the proposed membrane formulation. 
Fig.~\ref{f:cushion_contact:defo} shows the deformation of balloon and cushion for various gravity levels.
As shown, the deformation becomes very large, which makes the problem very challenging. The example is interesting as it involves large deformations, contact, pressure loading, hydrostatic loading and two volume constraints.



\subsection{Growth of a hemispherical water droplet}\label{s:droplet}

\begin{figure}[h]
\begin{center} \unitlength1cm
\includegraphics[height=59mm]{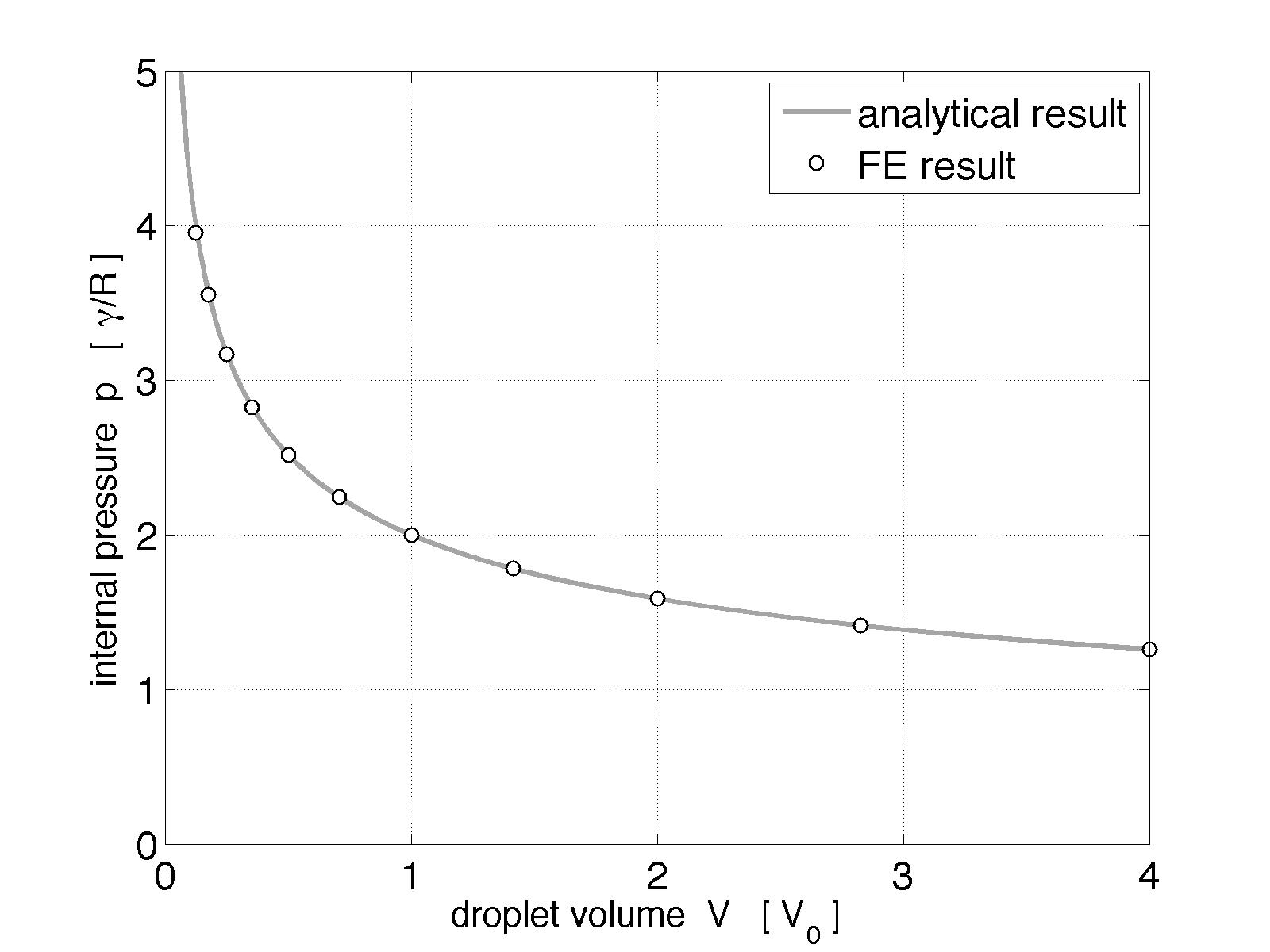}
\includegraphics[height=59mm]{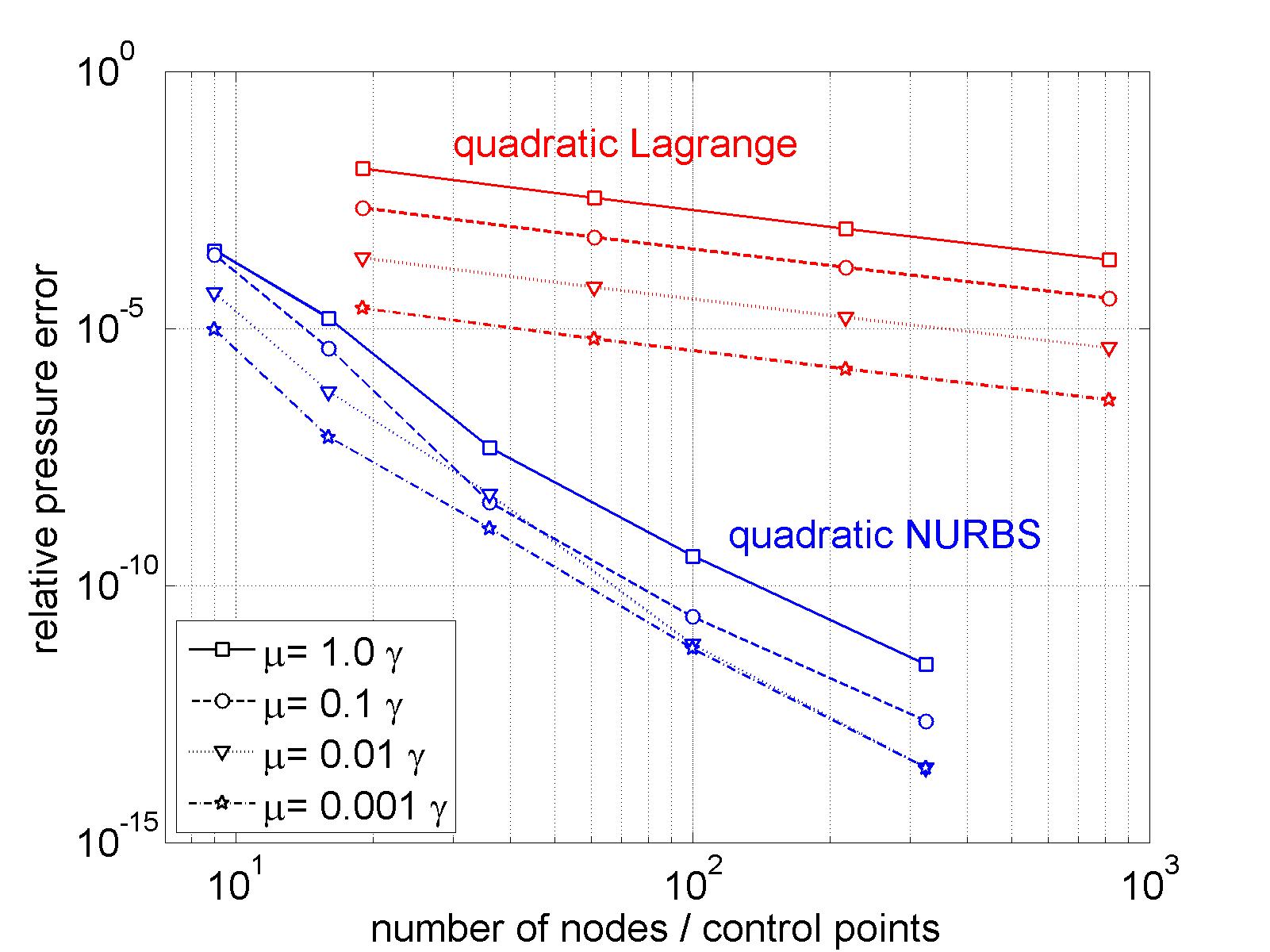}
\put(-15.7,0){a.}
\put(-8.0,0){b.}
\caption{Growing liquid droplet: (a) pressure-volume relation for $V\in[1/8~4]V_0$ (FE result for 12 quadratic FE with $\mu=0.01\gamma$); (b) convergence behavior for $V=4\,V_0$ and various $\mu$, considering $3\times3$ Gaussian quadrature points.}
\label{f:droplet_growth:conv}
\end{center}
\end{figure}

As a validation of the formulation for liquid membranes, we consider the growth of a hemispherical droplet resting on a rigid substrate and maintaining a contact angle of $90^\circ$. 
The problem is similar to the balloon inflation example (Sec.~\ref{s:balloon}), and the same FE meshes are used.
For a liquid water membrane the membrane stress is given by Eq.~(\ref{e:water}), i.e. the stress is deformation independent. This implies that only the out-of-plane but not the in-plane forces provide stiffness and the formulation is unstable in itself. The formulation can be stabilized by adding deformation dependent in-plane forces through Eq.~(\ref{e:fintio}.1).
These forces should not influence the out-of-plane behavior such that the original liquid membrane formulation remains unaffected.
We simply use the incompressible Neo-Hookean model to provide the additional in-plane stability. The Neo-Hookean material parameter $\mu$ then becomes a numerical stability parameter that should not affect the physical results. The internal forces acting on the finite element nodes are then simply given by
\eqb{l}
\mf_\mathrm{int}^e = \mf_\mathrm{int}^e\big(\sig^{\alpha\beta}_\mathrm{liquid}\big) + \mf_\mathrm{inti}^e\big(\sig^{\alpha\beta}_\mathrm{solid}\big)~.
\label{e:split}\eqe
For this example, like in Sec.~\ref{s:balloon}, the pressure-volume relation is also know analytically.
Setting $\sig = pr/2 = \gamma$, with $r=\lambda R$ and $V=\lambda^3V_0$, we find
\eqb{l}
\ds\frac{pR}{\gamma} = 2\Big(\frac{V_0}{V}\Big)^\frac{1}{3}~.
\eqe
Since the pressure remains positive (and is thus stabilizing the structure) the volume can also be decreased.
The computed pressure and the convergence of the proposed finite element formulation to the analytical result are shown in Fig.~\ref{f:droplet_growth:conv}.
Several values for the numerical stability parameter $\mu$ are considered. They all converge to the desired analytical result. Quadratic Lagrange and NURBS elements are considered, and it is seen that the NURBS formulation converges much faster. This is attributed to the higher surface continuity that appears in $\mf_\mathrm{inti}^e$ according to Eq.~(\ref{e:fintio}.1).
Decreasing $\mu$ improves the accuracy.
A more detailed analysis of the model proposed in Eq~(\ref{e:split}) along with the effect of parameter $\mu$ is required and will be considered in the future.

\subsection{Liquid droplet on a rigid substrate}\label{s:wcontact}

\begin{figure}[h]
\begin{center} \unitlength1cm
\includegraphics[height=44mm]{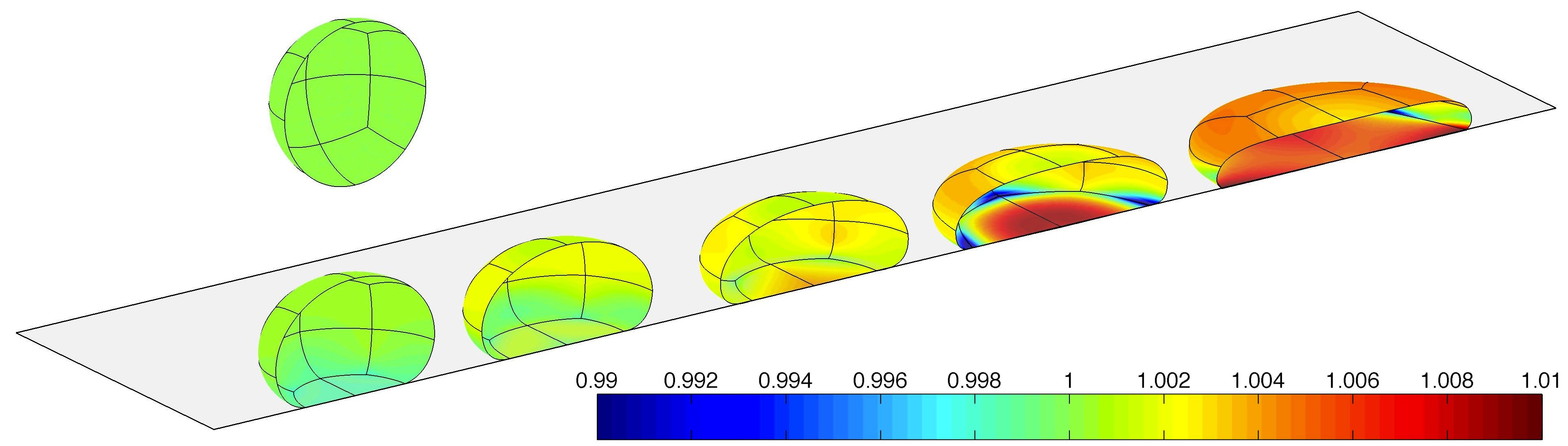}
\put(-15.7,0){a.}
\\[8mm]
\includegraphics[height=44mm]{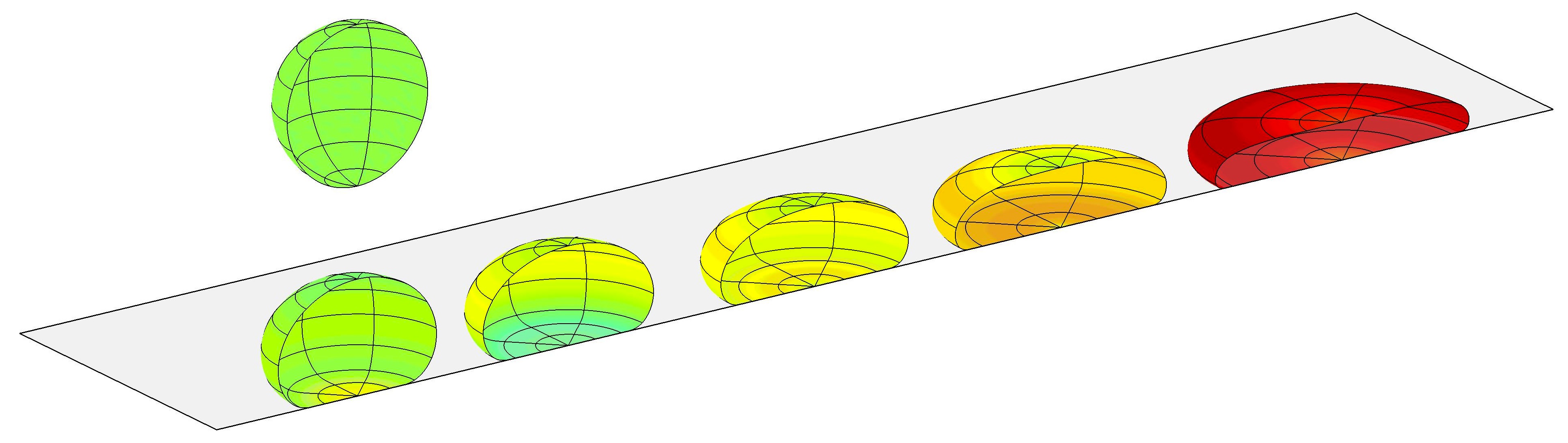}
\put(-15.7,0){b.}
\caption{Liquid droplet contact: droplet deformation for $\rho g=\{1,\,2,\,4,\,8,\,20\}\gamma/R^2$, each with identical volume and $\mu=0.005\,\gamma$: (a) quadratic Lagrange elements, (b) cubic T-spline elements. The color shows the membrane stress $I_1/2$ normalized by the surface tension $\gamma$. In theory $I_1/(2\gamma)=1$.}
\label{f:droplet_contact:defo}
\end{center}
\end{figure}

The last example examines a static water droplet in contact with a rigid substrate.
A distinct feature of liquid membranes is that they can form sharp contact angles at the contact boundary. In this case, the membrane surface forms a kink at the contact boundary. These surface discontinuities are associated with out-of-plane line forces. Such forces are not considered in the present framework. We thus consider a contact angle of $180^\circ$.
Initially, prior to loading and contact, the droplet is spherical. We denote the initial radius $R$, and the initial volume $V_0:=4\pi R^3/3$. The water inside the droplet is considered incompressible such that the volume remains constant during deformation. The weight of the water causes hydrostatic pressure loading of the membrane leading to contact with the substrate. Fig.~\ref{f:droplet_contact:defo} shows the computed droplet deformation for various gravity values considering different FE formulations. 
As the gravity level increases the droplet spreads out on the substrate. Formulation~(\ref{e:split}) is used for the computations. Stability parameter $\mu$ is set to 0.005\,$\gamma$.
For this $\mu$, the errors in the surface tension are less than 2.5\% in the case of quadratic Lagrange elements (Fig~\ref{f:droplet_contact:defo}, top.) and less than 1\% in the case of cubic T-splines (Fig~\ref{f:droplet_contact:defo}, bottom). In both cases the errors increase along with $\rho g$.
In the case of Lagrange elements the largest errors are found at the element boundaries, where the formulation is only $C^0$-continous. For the T-spline case, which is $C^2$ continous over the entire surface (appart from two degenerate points at the top and bottom), the error is uniformly spread over the surface.
It is remarked that the presented droplet model is much more general than the classical FE droplet formulation of \citet{brown80b}. This is discussed in detail in a forthcoming publication.


\section{Conclusion}

A novel computational formulation that is suitable for both solid and liquid, i.e. surface-tension-driven, membranes is presented. The theory, outlined in Sec.~\ref{s:membrane}, is based on the differential geometry of curved surfaces, allowing for a very general formulation that accounts for large deformations and general material laws. Curvilinear coordinates are used to formulate the surface geometry, kinematics, constitution, and balance laws.
The governing strong and weak forms are split into the in-plane and out-of-plane parts, allowing the use of different approximation techniques for both parts and an elegant treatment of liquid membranes. Also, the consideration of deformation-dependent pressure loading comes naturally within the proposed formulation. Various constraints imposed upon the membranes can be handled by the theory, including volume, area, and contact constraints.\\
The membrane formulation is discretized using nonlinear finite elements. This results in a very efficient formulation that only uses three degrees of freedom per surface node and avoids the use of local cartesian coordinate systems and the transformation of derivatives. This is discussed in Sec.~\ref{s:fed}. \\
The capabilities of the formulation are demonstrated by several challenging examples in Sec.~\ref{s:ne}. Linear Lagrange, quadratic Lagrange, quadratic NURBS, and cubic T-spline finite elements are considered for the discretization. Constraints are imposed using the Lagrange multiplier method in the case of volume constraints and the penalty method for contact constraints. The inflation of a balloon and the growth of a droplet are used to validate the solid and liquid membrane formulations and they both yield excellent results. Comparing the different finite element types, the examples show that large accuracy gains lie between linear and quadratic Lagrange, and between quadratic Lagrange and isogeometric finite elements.

The presented membrane formulation has been successfully applied to liquid droplets in this paper, but a rigorous analysis is still needed to assess the approach proposed in Eq.~(\ref{e:split}).
Another important extension to the present formulation is the inclusion of bending stiffness, which can be present in both fluid and solid films. In the latter case this should lead naturally to a rotation-free shell formulation, which can be suitably handled by isogeometric finite elements.
A further interesting extension is the consideration and development of different membrane material laws. Such a development is especially important for the case of biological membranes, which are often characterized by complex material behavior.


\appendix

\section{Consistent linearization of various quantities}\label{s:lin}

For Newton's method we need to linearize the kinematical quantities of the discrete system at $\mx$ in the direction $\Delta\mx$. This is done at the FE level.

\subsection{Linearization of $\ba_\alpha$}

According to Eq.~(\ref{e:aapprox}) we have
\eqb{l}
\Delta\ba_{\alpha} = \mN_{,\alpha}\,\Delta\mx_e~.
\label{e:Dba}
\eqe

\subsection{Linearization of $a_{\alpha\beta}$}

With definition~(\ref{e:aab}) follows
\eqb{ll}
\label{eq:linaAlpBeta}
\Delta a_{\alpha\beta}= \big(\ba_{\alpha}\cdot\mN_{,\beta} + \ba_{\beta}\cdot\mN_{,\alpha}\big)\Delta\mx_e~.
\eqe

\subsection{Linearization of $J$}

The change $\Delta J$ can be written as
\eqb{l}
\Delta J = \ds\pa{J}{\ba_{\alpha}}\cdot \Delta\ba_\alpha~,
\eqe
where 
\eqb{l}
\ds\pa{J}{\ba_{\alpha}}=J\ba^{\alpha}~.
\eqe
Thus
\eqb{l}
\Delta J = J\ba^{\alpha}\cdot\mN_{,\alpha}\,\Delta\mx_e~.
\eqe

\subsection{Linearization of $a^{\alpha\beta}$}

From Eq.~(\ref{e:aa}) and the formula
\eqb{l}
a^{\alpha\beta} = \ds\frac{1}{a}e^{\alpha\gamma\,}a_{\gamma\delta}\,e^{\delta\beta}~,\quad
a := \det a_{\alpha\beta}~,
\eqe
where 
\eqb{l}
\big[e^{\alpha\beta}\big] = \left[\begin{array}{cc}
        0 & 1 \\
         -1 & 0
        \end{array}\right]
\eqe
is the unit alternator, we find
\eqb{ll}
\Delta a^{\alpha\beta}= m^{\alpha\beta\gamma\delta}\,\ba_{\gamma}\cdot\mN_{,\delta} \,\Delta\mx_e~,
\eqe
with
\eqb{l}
        m^{\alpha\beta\gamma\delta} = \ds\frac{1}{a}\big(e^{\alpha\gamma}e^{\beta\delta} + e^{\alpha\delta}e^{\beta\gamma}\big)-  2a^{\alpha\beta}a^{\gamma\delta}~.
\eqe

\subsection{Linearization of $\bn\,\dif a$}

The surface normal $\bn$ appears together with the area element $\dif a$ and it is convenient to linearize them together. According to Eqs.~(\ref{e:n}) and (\ref{e:da}) we have
\eqb{l}
\bn\,\dif a = \ba_1\times\ba_2\,\dif\Box~.
\eqe
Hence
\eqb{l}
\Delta(\bn\,\dif a) = 
\ds\sum_I\Big(N_{,1}\,\Delta\bx_I\times\ba_2 + \ba_1\times N_{,2}\,\Delta\bx_I\Big)\,\dif\Box~.
\eqe
Expanding $\Delta\bx_I$ into $\Delta\bx_I=\Delta x_I^\alpha\,\ba_\alpha + \Delta x_I^\mrn\,\bn$ we find
\eqb{lllll}
\Delta\bx_I\times\ba_2 \is J_a\big(\bn\,\Delta x_I^1 - \ba^1\,\Delta x_I^\mrn \big) 
	\is J_a\big(\bn\otimes\ba^1-\ba^1\otimes\bn\big)\,\Delta\bx_I~, \\[2mm]
\ba_1\times\Delta\bx_I \is J_a\big(\bn\,\Delta x_I^2 - \ba^2\,\Delta x_I^\mrn \big)
	\is J_a\big(\bn\otimes\ba^2-\ba^2\otimes\bn\big)\,\Delta\bx_I~,
\eqe
where $J_a=\sqrt{\det a_{\alpha\beta}} = \dif a/\dif\Box$. Thus
\eqb{l}
\Delta(\bn\,\dif a) = \big(\bn\otimes\ba^\alpha-\ba^\alpha\otimes\bn\big)\,\mN_{,\alpha}\,\Delta\mx_e\,\dif a~.
\label{e:Dnda}\eqe

\subsection{Linearization of $\tau^{\alpha\beta}$}

For the solid model according to Eq.~(\ref{e:incompNH_ab}) we have
\eqb{l}
\Delta\tau^{\alpha\beta} = \mu T\big(2 J^{-3}\,\Delta J\, a^{\alpha\beta} -J^{-2}\,\Delta a^{\alpha\beta}\big)~,
\eqe
which can be rewritten into
\eqb{l}
\Delta\tau^{\alpha\beta} = c^{\alpha\beta\gamma\delta}\,\ba_{\gamma}\cdot\mN_{,\delta}\,\Delta\mx_e~,
\label{e:lintau}\eqe
with
\eqb{l}
\ds c^{\alpha\beta\gamma\delta}=\mu TJ^{-2}\Big(4a^{\alpha\beta}a^{\gamma\delta} - \frac{1}{a} \big(e^{\alpha\gamma}e^{\beta\delta} + e^{\alpha\delta}e^{\beta\gamma}\big)\Big)~.
\eqe
Note that the tensor $[c^{\alpha\beta\gamma\beta}]$, like $[m^{\alpha\beta\gamma\beta}]$, posses both major and minor symmetries.\\
For the liquid model according to Eq.~(\ref{e:water}) we have
\eqb{l}
\Delta\tau^{\alpha\beta} = \gamma\big(\Delta J\, a^{\alpha\beta}  + J\,\Delta a^{\alpha\beta}\big)~,
\eqe
which can also be written in the form~(\ref{e:lintau}), where now
\eqb{l}
\ds c^{\alpha\beta\gamma\delta}=\gamma J\Big(\frac{1}{a} \big(e^{\alpha\gamma}e^{\beta\delta} + e^{\alpha\delta}e^{\beta\gamma}\big) - a^{\alpha\beta}a^{\gamma\delta}\Big)~.
\eqe

\section{Finite element tangent matrices}\label{s:LFE}

\subsection{Tangent matrix associated with $G^e_\mathrm{int}$}

The internal force vector $\mf^e_\mathrm{int}$, given in Eq.~(\ref{e:fint}), yields
\eqb{l}
\Delta\mf_\mathrm{int}^e 
= \ds\int_{\Omega^e_0}\mN^T_{,\alpha}\,\Delta\tau^{\alpha\beta}\,\mN_{,\beta}\,\dif A\,\mx_e
+ \ds\int_{\Omega^e_0}\mN^T_{,\alpha}\,\tau^{\alpha\beta}\,\mN_{,\beta}\,\dif A\,\Delta\mx_e~.
\label{e:Dfint}\eqe
In view of Eq.~(\ref{e:lintau}), we can write
\eqb{l}
\Delta\mf^e_{\mathrm{int}}=\big(\mk^e_{\mathrm{mat}} + \mk^e_{\mathrm{geo}}\big)\,\Delta \mx_e
\eqe
where we have introduced the material stiffness matrix
\eqb{l}
\mk^e_{\mathrm{mat}} = \ds\int_{\Omega^e_0}c^{\alpha\beta\gamma\delta}\,\mN^T_{,\alpha}\, (\ba_{\beta}\otimes\ba_{\gamma})\,\mN_{,\delta}\, \dif A
\eqe
and the geometric stiffness matrix
\eqb{l}
\mk^e_{\mathrm{geo}} = \ds\int_{\Omega^e_0}\mN_{,\alpha}^T\,\tau^{\alpha\beta}\,\mN_{,\beta}\,\dif A
\eqe
Both these matrices are symmetric for the two constitutive models considered here. For those models, the terms in
$\mk^e_\mathrm{mat}$ should be multiplied-out a-priory to obtain an efficient implementation. 
If we consider splitting $\mf_\mathrm{int}^e$ into $\mf_\mathrm{inti}^e$ and $\mf_\mathrm{into}^e$ additional stiffness terms are picked up. These are reported in a forthcoming publication.

\subsection{Tangent matrix associated with $G^e_\mathrm{ext}$}

From Eq.~(\ref{e:fext}), for dead $\bff_0$ and $\bar\bt$, we have
\eqb{l}
\Delta\mf_\mathrm{ext}^e = \ds\int_{\Omega^e} \mN^T\,\bn\,\Delta p\,\dif a
+\int_{\Omega^e} \mN^T\,p\,\Delta(\bn\,\dif a)~.
\eqe
The first term is only required for hydrostatic loading according to Eq.~(\ref{e:ph}). Here we find
\eqb{l}
\Delta p = -\rho\,\bg\,\mN\,\Delta\mx_e~.
\eqe
Contribution $\Delta(\bn\,\dif a)$ is given by Eq.~(\ref{e:Dnda}). As a result,
\eqb{l}
\mk_\mathrm{ext}^e = -\ds\int_{\Omega^e} \rho\,\mN^T\,\bn\otimes\bg\,\mN\,\dif a
+\int_{\Omega^e} p\,\mN^T\,\big(\bn\otimes\ba^\alpha-\ba^\alpha\otimes\bn\big)\,\mN_{,\alpha} \,\dif a~.
\eqe

\subsection{Tangent contributions associated with the volume constraint}

If the volume constraint $g_\mrv=0$ is active, we need to account for the unknown Lagrange multiplier $p_\mrv$ in the linearization.
For the external forces we now have
\eqb{l}
\Delta\mf^e_\mathrm{ext} = \mk^e_\mathrm{ext}\,\Delta\mx_e + \ml^e_\mathrm{ext}\,\Delta p_\mrv~,
\eqe
with
\eqb{l}
\ml_\mathrm{ext}^e = \ds\pa{\mf^e_\mathrm{ext}}{p_\mrv} =  \int_{\Omega^e} \mN^T\,\bn\,\dif a~.
\eqe
Further, at the element level,
\eqb{l}
\Delta g^e_\mrv = \mh^e_\mrv\,\Delta\mx_e~,
\eqe
with
\eqb{l}
\mh_\mrv^e = \ds\pa{g^e_\mrv}{\mx_e} =  \frac{1}{3}\int_{\Omega^e} \bn\cdot\mN\,\dif a
+  \frac{1}{3}\int_{\Omega^e} \bx\cdot\big(\bn\otimes\ba^\alpha-\ba^\alpha\otimes\bn\big)\,\mN_{,\alpha}\,\dif a~.
\eqe
The preceding contributions can be arranged into the elemental tangent matrix
\eqb{l}
\mk^e :=\left[\begin{array}{cc}
       \mk^e_\mathrm{int}-\mk^e_\mathrm{ext} ~ & -\ml^e_\mathrm{ext} \\[6mm]
       \mh^e_\mrv & 0
\end{array}\right],
\eqe
which describes the change in $\mf_\mathrm{int}-\mf_\mathrm{ext}$ and $g_\mrv$ due to changes in position $\mx_e$ and pressure $p_\mrv$.


\bigskip

{\Large{\bf Acknowledgements}}

The authors are grateful to the German Research Foundation (DFG)
for supporting this research under projects SA1822/3-2, SA1822/5-1 and GSC 111.

\bigskip

%

\bibliographystyle{apalike}
\bibliography{bibliography}

\end{document}